\documentclass[fleqn,usenatbib]{mnras}

\usepackage{newtxtext,newtxmath}
\usepackage[T1]{fontenc}

\DeclareRobustCommand{\VAN}[3]{#2}
\let\VANthebibliography\thebibliography
\def\thebibliography{\DeclareRobustCommand{\VAN}[3]{##3}\VANthebibliography}

\usepackage{graphicx}	
\usepackage{amsmath}	
\usepackage{epstopdf}
\usepackage{multicol}   
\usepackage{bm}
\usepackage{color}
\usepackage{xspace}
\usepackage{siunitx}
\usepackage{threeparttable}

\usepackage{CJKutf8}

\usepackage{bm}
\usepackage{color}
\usepackage{ulem}
\usepackage{xspace}

\definecolor{mygray}{gray}{0.6}
\definecolor{orange}{rgb}{1.0, 0.4, 0.0}
\definecolor{myblue}{rgb}{0.1, 0.5, 0.7}
\definecolor{TsinghuaPurple}{cmyk}{0.58,0.90,0,0}
\definecolor{magenta}{rgb}{0.858, 0.188, 0.478}
\newcommand{\ccc}[1]{\textcolor{orange}{[\textit{chris: \small #1}]}}

\newcommand{\corem}[1]{\textcolor{mygray}{\sout{#1}}}
\newcommand{\yuc}[1]{\textcolor[RGB]{139,0,139}{[\textit{\small #1}]}}

\newcommand{\cim}[1]{\textcolor{orange}{[\textit{$\star$chris$\star$}:\textbf{\small #1}}]}

\newcommand{\app}[1]{App.~\ref{sec:#1}}
\newcommand{\Se}[1]{Section~\ref{sec:#1}}
\newcommand{\se}[1]{Sect.~\ref{sec:#1}}
\newcommand{\Fg}[1]{Figure~\ref{fig:#1}}
\newcommand{\fg}[1]{Fig.~\ref{fig:#1}}
\newcommand{\fgs}[2]{Figs.~\ref{fig:#1} and \ref{fig:#2}}
\newcommand{\tb}[1]{Table~\ref{tab:#1}}

\newcommand{\eq}[1]{Eq.~(\ref{eq:#1})}
\newcommand{\Eq}[1]{Equation~(\ref{eq:#1})}

\newcommand{\eqs}[2]{Eqs.~(\ref{eq:#1}) and (\ref{eq:#2})}
\newcommand{\eqto}[2]{Eqs.~(\ref{eq:#1})--(\ref{eq:#2})}

\newcommand\gas{\mathrm{g}}

\newcommand\parti{\mathrm{p}}
\newcommand\tracer{\mathrm{tr}}

\usepackage{ulem}
\makeatletter
\def\uwave{\bgroup \markoverwith{\lower3.5\p@\hbox{\sixly \textcolor{red}{\char58}}}\ULon}
\font\sixly=lasy6 
\makeatother

\usepackage{orcidlink}

\title[Pebble Recycling]{Atmospheric Recyling of Volatiles by Pebble-Accreting Planets}

\author[Wang et al.]{
Yu Wang (\begin{CJK*}{UTF8}{gbsn}王雨\end{CJK*})$^{1}$ \orcidlink{0009-0004-2217-4439},
Chris W. Ormel $^{1}$ \thanks{E-mail: chrisormel@tsinghua.edu.cn} \orcidlink{0000-0003-4672-8411},
Pinghui Huang (\begin{CJK*}{UTF8}{gbsn}黄平辉\end{CJK*})$^{2,3}$ \orcidlink{0000-0002-7575-3176}, Rolf Kuiper$^{4}$ \orcidlink{0000-0003-2309-8963}
\\
$^{1}$Department of Astronomy, Tsinghua University, 30 Shuangqing Rd, Haidian DS 100084, Beijing, China\\
$^{2}$University of Victoria, Victoria, British Columbia, Canada\\
$^{3}$Institute for Advanced Study, Tsinghua University, 30 Shuangqing Rd, Haidian DS 100084, Beijing, China\\
$^{4}$Faculty of Physics, University of Duisburg-Essen, Lotharstra{\ss}e 1, 47057 Duisburg, Germany
}

\date{Accepted XXX. Received YYY; in original form ZZZ}

\pubyear{2023}

\begin{document}
\label{firstpage}
\pagerange{\pageref{firstpage}--\pageref{lastpage}}
\maketitle

\begin{abstract}
    Planets, embedded in their natal discs, harbour hot envelopes. When pebbles are accreted by these planets, the contained volatile components may sublimate, enriching the envelope and potentially changing its thermodynamical properties.
    However, the envelopes of embedded planets actively exchange material with the disc, which would limit the buildup of a vapour-rich atmosphere.
    To properly investigate these processes, we have developed a new phase change module to treat the sublimation process with hydrodynamical simultions. Combined with the recently developed multi-dust fluid approach, we conduct 2D self-consistent hydrodynamic simulations to study how pebble sublimation influences the water content of super-Earths and sub-Neptunes.
    We find the extent and the amount of vapour that a planet is able to hold on to is determined by the relative size of the sublimation front and the atmosphere. When the sublimation front lies far inside the atmosphere, vapour tends to be locked deep in the atmosphere and keeps accumulating through a positive feedback mechanism. On the other hand, when the sublimation front exceeds the (bound) atmosphere, the ice component of incoming pebbles can be fully recycled and the vapour content reaches a low, steady value. Low disc temperature, small planet mass and high pebble flux (omitting accretion heating by pebbles) render the planet atmosphere vapour-rich while the reverse changes render it vapour-poor.  
    The phase change module introduced here can in future studies also be employed to model the chemical composition of the gas in the vicinity of accreting planets and around snowlines.
\end{abstract}



\section{Introduction}
Planets are born in gaseous discs. The close-in, low-mass planet population revealed by the \textit{Kepler} and the Transiting Exoplanet Survey Satellite (\textit{TESS}) missions, known as super-Earths and sub-Neptunes \citep{FressinEtal2013,WinnFabrycky2015,ZhuEtal2018,ZhuDong2021}, are believed to hold -- or have held in the past -- moderate amounts of H and He gas. This strongly suggests these planets have their genesis in the gas-rich disc \citep{FultonEtal2017,OwenWu2017,JinMordasini2018}. 

In the core accretion model, planets need to grow massive enough before they start to bind an atmosphere \citep{Mizuno1980,PollackEtal1996}.  Low-mass planets can be defined to have their atmosphere in hydrostatic balance with the disc. Their imprints on discs remain limited, in contrast to high-mass planets which open gaps and that have been proposed to be responsible for the substructure seen in young discs \citep{BenistyEtal2017,AndrewsEtal2018,LongEtal2018, ZhangEtal2018,AvenhausEtal2018, vanderMarelEtal2019}.
To understand the transition from low-mass planets to gas giants, one must understand how efficient atmospheres cool \citep{LeeEtal2014,GinzburgSari2017,Yu2017,OrmelEtal2021}. For the post-disc phase, atmospheres are further affected by the stellar insolation \citep{LopezFortney2013,OwenWu2017,JinMordasini2018} or the heat release from the core \citep{GinzburgEtal2018,VazanEtal2018,VazanEtal2018i,GuptaSchlichting2019}. Importantly, all these works consider a fixed composition for the gas (the standard H and He-dominated gas) and also assume that the atmospheres stay thermodynamically isolated in the disc phase.

These assumptions can be challenged, in particular when planet assembly occurs by accreting pebbles \citep{OrmelKlahr2010,LambrechtsJohansen2012}.
As atmospheres around low-mass planets heat up, the more volatile material component contained in the pebbles sublimates \citep{Alibert2017,BrouwersEtal2018}. Pebble sublimation carries major implications. First, the atmosphere will be polluted by the high-Z vapour and has a larger mean molecular weight \citep{IaroslavitzPodolak2007,BodenheimerEtal2018}. The polluted atmosphere becomes significantly heavier than the pure H-He atmosphere, thus triggering runaway gas accretion at a lower mass \citep{Wuchterl1993,VenturiniEtal2015,VenturiniEtal2016}, which is renamed the `critical metal mass' by \citet{BrouwersOrmel2020}. 

Second, pebble sublimation in atmosphere can affect the volatile delivery by setting an effective iceline inside the planet atmosphere.
This contrasts the standard assumption that the chemical inventory of a planet atmosphere is set by the local disc properties.  That is, a planet accretes solids (ices) if it is situated outside the \textit{disc} iceline, while it accretes vapour only interior to it \citep{OebergEtal2011,SchoonenbergOrmel2017,DrazkowskaAlibert2017,IdaEtal2019,BoothIlee2019,KrijtEtal2022,Molli`ereEtal2022}.
The super-solar $N_2$ abundance of Jupiter, for example, has been suggested as an imprint for its core to be assembled outside the $\mathrm{N_{2}}$ snowline \citep{OebergWordsworth2019,BosmanEtal2019}. 
However, \citet{BarnettCiesla2022} pointed out that an accreting planet would greatly elevate the ambient temperature and halt $\mathrm{N_{2}}$ delivery by sublimation, putting even more stringent constraints on Jupiter's core birth locations. 
Similarly, \citet{JohansenEtal2021} proposed that Earth-like planets' atmosphere can be largely refreshed by the deep recycling flows from the disc, which potentially remove the sublimated vapour and prevent Earth from becoming abundant in water and carbon, even though the respective disc snowlines move interior to the Earth's location \citep{JohansenEtal2023,JohansenEtal2023i,JohansenEtal2023ii}. Knowing the efficacy of the recycling process is therefore instrumental to understand the composition of planets and to guide characterization efforts of exoplanets.

The recycling mechanism has been studied with 2D and 3D hydrodynamical simulations, which solve for the flow patterns in the vicinity of small planets embedded in the disc. In 2D, the gas comes in and leaves through the horseshoe orbit and there is an inner bound region supported by rotation \citep{OrmelEtal2015,FungEtal2015,BethuneRafikov2019}. On the other hand, in 3D gas tends to be accreted from the polar direction and flows back to the disc in the equatorial plane without a clear boundary between the disc and the atmosphere \citep{OrmelEtal2015i,CimermanEtal2017,LambrechtsLega2017,KurokawaTanigawa2018,PopovasEtal2018,FungEtal2019,KuwaharaEtal2019,MoldenhauerEtal2021,MoldenhauerEtal2022}. Accounting for radiation transport in 3D, \citet{MoldenhauerEtal2021} finds a fully-recycled atmosphere for an Earth-mass planet, which halts cooling and subsequent (runaway) gas accretion by continuously refreshing the low-entropy atmosphere with the higher entropy gas from the disc. Still, these simulations assume a single component H-He gas and it is unclear whether the recycling mechanism also operates in polluted atmosphere.  That is, what happens to the (volatile) components that pebbles contain: do they stay in the atmosphere or do they flow back to the disc (recycle)? 

To investigate the fate of pebble sublimation for low-mass planets (super-Earths), we have designed a new phase change module for Athena++, on top of the recently developed multi-dust fluid module by \citet{HuangBai2022}. The phase change module treats the mass transfer, energy exchange and momentum conservation processes during sublimation and condensation (freezing-out) self-consistently (\citealt{LiChen2019}), enabling us to appropriately study the coupled thermodynamic and hydrodynamical effect of phase change processes in numerical simulations for the first time. 
In this pioneering work, our focus lies on describing phase change processes: the sublimation of the volatile components of pebbles upon entering hot atmospheres, the thermodynamical consequences of the released vapour, and the freeze-out of vapour in the form of ice grains, which can flow back to the disc. Accordingly, we include only the minimum amount of physical processes (e.g., adiabatic equation of state, single ice species), omitting for the moment processes as feedback of pebbles on gas, grain growth, deposition of vapour on incoming pebbles, as well as radiation transport.  To further reduce the simulation runtime and explore the parameter space, simulations are carried out in 2D. Our over-arching goal is to identify trends of the pebble recycling process with changing disc conditions and planet mass -- trends that can be verified in the future with more detailed approaches. Simulation caveats are addressed in \se{caveats}. 

In this work, we focus on a single ice species -- water. We identify a dichotomy in the amount of water vapour a planet can hold on to. For recycling-dominated planets, the ice component of the incoming pebbles is fully recycled and the vapour content of the atmosphere reaches a steady, limited value. On the other hand, in the vapour-dominated regime, sublimation occurs deep in the atmosphere, out of reach for recycling. A few simulations, falling in between these limits, delineate these two outcomes. In general, low ambient disc temperatures, small planets, high pebble fluxes and large Stokes numbers push the planet towards the vapour-dominated limit. In \se{discussion} we compile the simulation results to obtain a predictive estimate of which regime applies to the combination of disc condition and pebble material properties -- an estimator that is imprecise in light of the above-mentioned limitations, but that we expect to convey the trends well.

The structure of the paper is as follows. In \se{model} we describe the basic work principles of the phase change module and our implementation in different hydrodynamic codes, which we conduct a few tests to verify the robustness of our model (see also \app{gas_only}). \Se{results} begins with the simulation set-up of our fiducial run and then introduce the main results of this paper: various flow patterns and vapour content gained in the atmosphere with different disc parameters and planet mass. We identifiy three qualitative regimes in this section and explain the underlying mechanisms. In \se{discussion} we first summarize the results and then propose an estimator to determine the evolving trend of vapour content. Then we discuss the caveats and future improvements and applications of our phase change module. \Se{conclusion} lists our conclusions.

\section{Model}
\label{sec:model}
This section describes the phase change extension of the multi-fluid module. In \se{design} we outline the problem and introduce our modelling approaches. In \se{hydro-eqs} we present the governing equations for both gas and the multi-particle fluid hydrodynamics, building upon the work of \citet{HuangBai2022}, and introduce our choices of parameters and source terms regarding the problem we focus on. In \se{phase change model} we introduce the phase change model with which we simulate evaporation fronts in hydrodynamical codes. Finally, in \se{implement}, we describe how we implement the model in both PLUTO and ATHENA++.

\begin{figure}
 \includegraphics[width=\columnwidth]{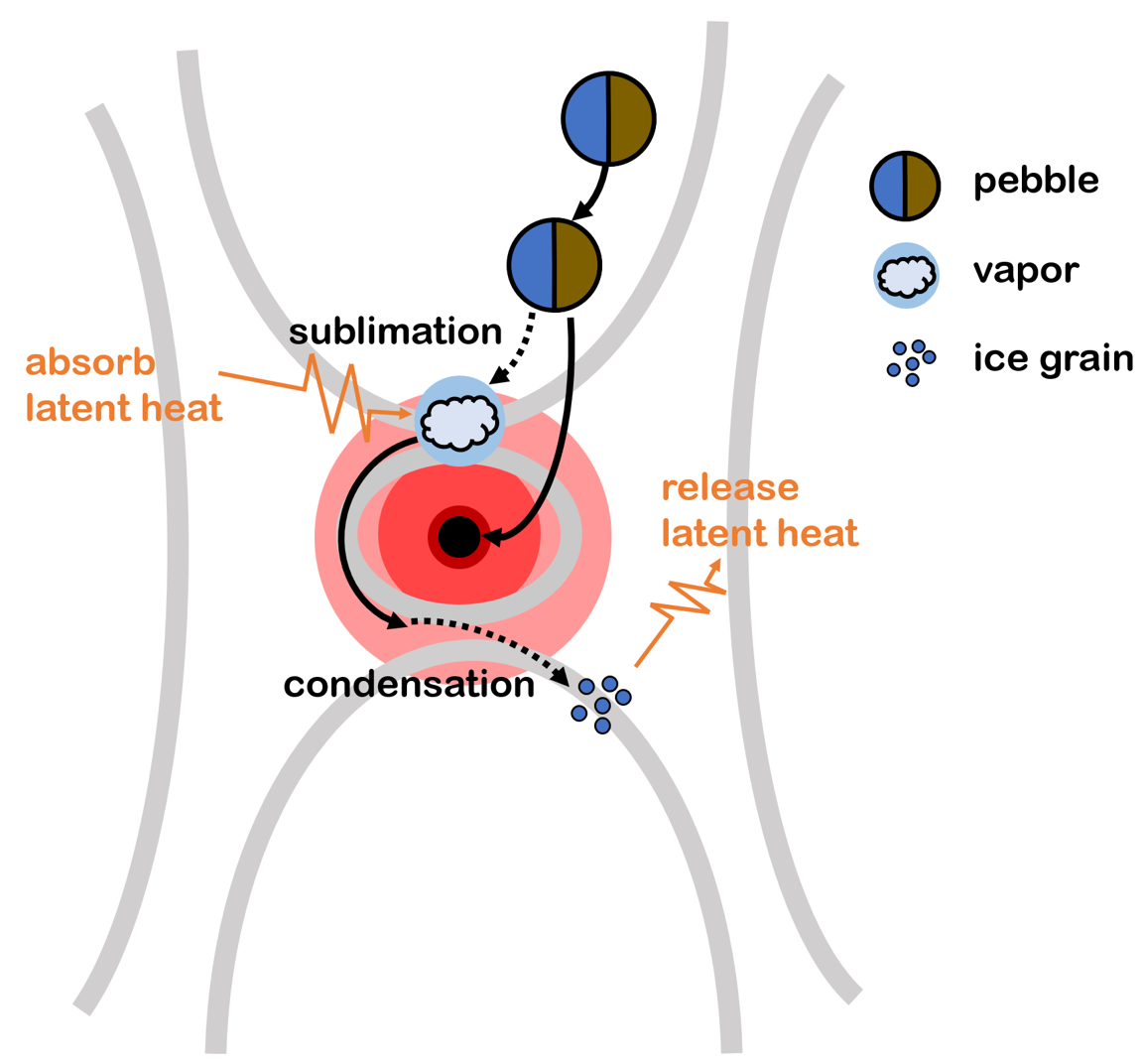}
 \caption{Schematic displaying the processes occurring at the sublimation fronts. Grey curves are gas streamlines as seen in the co-moving frame. Around the planetary core (black dot) a hot atmosphere forms. From top to bottom: pebbles consisting of ice and refractory materials approach the atmosphere of an embedded planet. In the hot atmosphere the ice sublimates off the pebbles, while the refractory remnant of the pebble proceeds towards the core.  vapour can recondense to ice grains in colder regions in the atmosphere, which may be advected back to the disc.}
 \label{fig:phase_change}
\end{figure}

\subsection{Overall model design}
\label{sec:design}
We illustrate the problem in \fg{phase_change}. In this figure, a planetary core (black dot at the centre) is embedded into the disc and surrounded by a denser atmosphere. Depicted in the co-moving frame, this results in the familiar horeshoe, atmosphere and circulating streamlines (grey lines; \citep{OrmelEtal2015,BethuneRafikov2019,MoldenhauerEtal2021}. In addition, there are pebbles, embedded in the flow, which are accreted by the planet (pebble accretion). When they approach the inner atmosphere, the temperature of the ambient gas is hot enough to sublimate the ice from the pebble, during which latent heat is absorbed from the environment. The released vapour can also flow back to the disc and recondense in the form of ice grains, which are carried away with the gas flow. On the other hand, the refractory components of the pebble continue their trajectory towards the planet surface. 

We simulate these processes with a multi-fluid modelling approach. In this setup, we distinguish between gas, tracer and particle fluids.
The ``gas fluid'' is the mixture of all non-condensable gaseous species (mostly H and He) and all vaporized ices, which are characterized by the standard gas properties as pressure, temperature and mean molecular weight. There is only one gas fluid. In addition, the density of each vapour species is followed by a ``tracer fluid''. The tracer fluid inherits the hydrodynamical properties (i.e., velocity field) of the gas fluid. Next, there are the particle fluids, with which the particle species are simulated. In general, each particle species is a compound consisting of various condensates and refractory materials. For example, cm-sized ``pebbles'' are a compound whose components are refractory silicates and water ice, while micron-sized water ice grains are another single-component compound. Each of these material components is described by a pressureless ``particle fluid" which inherits the aerodynamic properties of their parent compound. In \fg{multifluids}, we present an example of the multi-fluid setup. We assume there are two particle species A and B (varying in size e.g.) and they each contain one refractory component and one ice component, but in generally not in the same proportions. During the phase change, the ice component in the particles will be transferred to vapour, increasing the local concentration of vapour over non-condensible gas (here: hydrogen and helium). For the entire setup, we need four particle fluids to describe the refractory material and the ice of each particle compound, a single gas fluid for the H/He and vapour, and one tracer fluid to follow the vapour component corresponding to the single ice component.

\begin{figure}
    \includegraphics[width=\columnwidth]{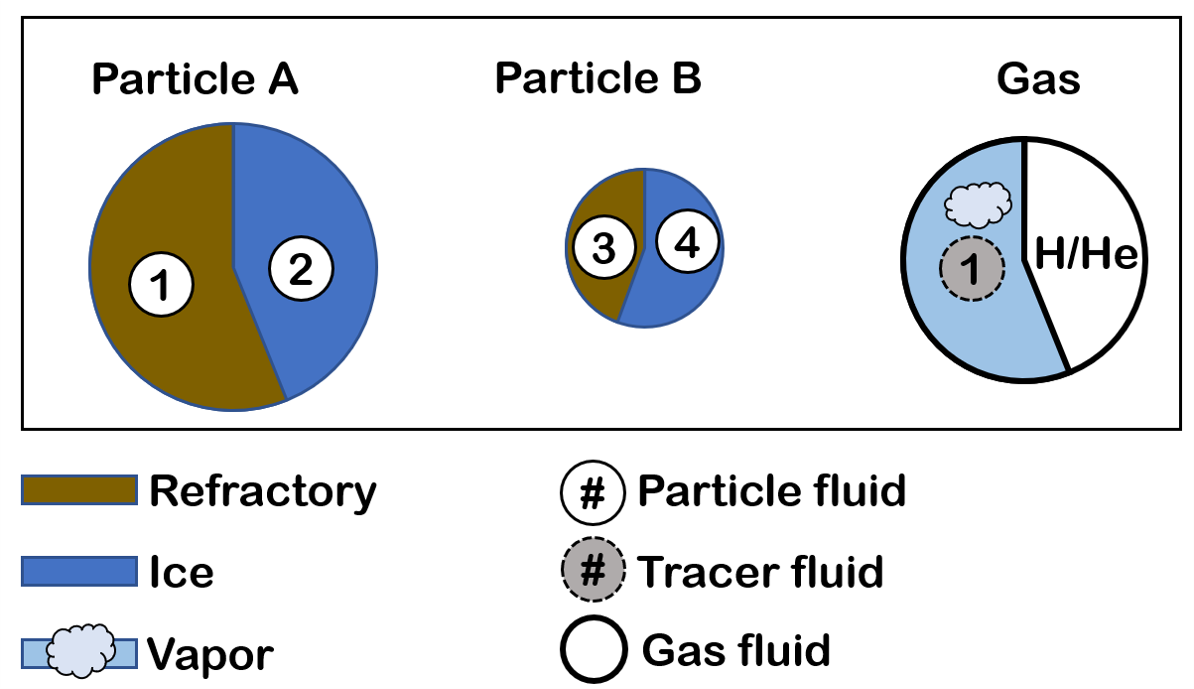}
    \caption{
       An example of the multi-fluid setup. There are two particle species A and B containing two material components: refractory and ice, which in total need four particle fluids to follow. The gas contains one non-condensable component (H, He mixture here) and one vapour component, which are described by one gas fluid in total and one additional tracer fluid.}
    \label{fig:multifluids}
\end{figure}

\begin{figure}
    \centering
    \includegraphics[width=\columnwidth]{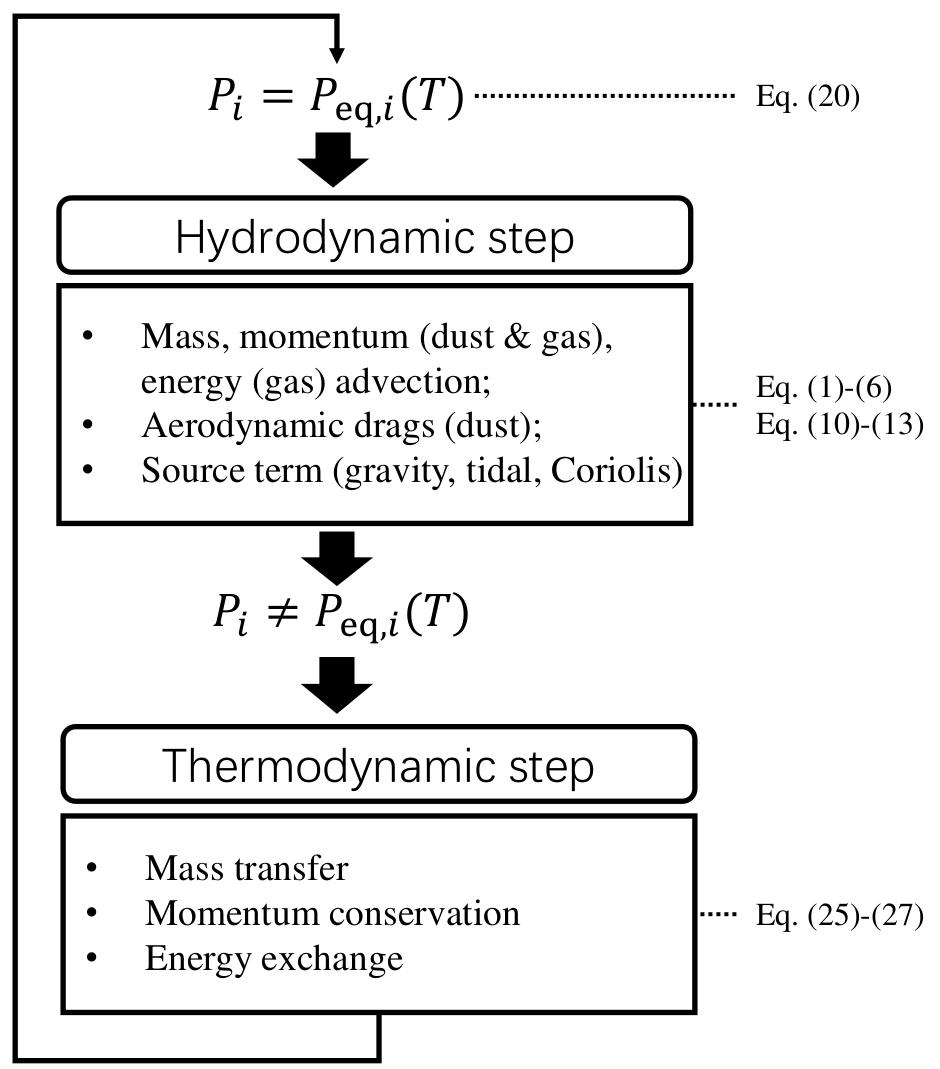}
    \caption{Flowchart representing the workflow of the simulation. It starts with the vapour partial pressure $P_{i}$ for material $i$ equaling the saturation pressure $P_{\mathrm{eq},i}$. And after the hydrodynamic step, the output doesn't satisfy it anymore. Then the thermodynamic step solves for $P_{i} = P_{\mathrm{eq},i}(T)$ with mass transfer, energy exchange and momentum conservation. On the right, we list the equations related to both the hydrodynamic and thermodynamic steps.}
    \label{fig:flowchart}
   \end{figure}

The simulation is separated into a hydrodynamic and a thermodynamic step. \Fg{flowchart} describes the workflow formally. Initially $P_{i}$, the pressure of vapour component $i$,  meets the vapour saturation pressure equation ($P_{i} = P_{\mathrm{eq},i}(T)$, \eq{saturated}) and serves as input of the simulation. The hydrodynamic step solves a set of multi-fluid hydrodynamical equations to update the density and velocities of the gas and particle. After this step, $P_{i} = P_{\mathrm{eq},i}(T)$ is no longer held. For example, the saturation pressure will exceed the partial pressure of vapour due to temperature increase, which leads to sublimation. Then, the thermodynamic step is needed to update the properties of the gas (temperature, density and vapour fraction) with the phase change model, which represents the major new addition of this work. This model solves the mass transfer, momentum and energy exchange between ice and gas as determined by the vapour saturation pressure equation. It connects the solid and gas phases of volatiles. As a result, we are able to investigate both the hydrodynamics and thermodynamics of volatiles in a self-consistent way. We will describe both steps in detail in the next two sections.

\subsection{Governing equations}
\label{sec:hydro-eqs}
The hydrodynamic equations of gas and particle fluids are presented in conservative form. 
We use subscripts ``g" ``p" and ``tr" to denote gas, particles and the tracers respectively. We also use ``z'' to denote the material components. Furthermore, let $N_{\parti}$ and $N_{\mathrm{z}}$ describe the total number of particle species and material components (both refractory and volatile). Thus, in total we have one gas fluid, $N_{\mathrm{z}} \times N_{\parti}$ particle fluids and at most $N_{\mathrm{z}}$ tracer fluids (there are no refractory tracer fluids.).

The governing equations read:

\begin{equation}
\label{eq:CE_gas}
    \frac{\partial \rho_{\gas}}{\partial t} + \nabla \cdot (\rho_{\gas} \bm{v}_{\gas}) = 0,
\end{equation}

\begin{equation}
\label{eq:ME_gas}
    \frac{\partial (\rho_{\gas} \bm{v}_{\gas})}{\partial t} + \nabla \cdot (\rho_{\gas} \bm{v}_{\gas} \bm{v}_{\gas} + P_{\gas} \bm{I}) = \rho_{\gas} \bm{f}_{\mathrm{g, src}},
\end{equation}

\begin{equation}
\label{eq:EE}
\begin{aligned}
\frac{\partial E_{\gas}}{\partial t}&+\nabla \cdot\left[\left(E_{\gas}+P_{\gas}\right) \bm{v}_{\gas} \right]  = \rho_{\gas} \bm{f}_{\gas, \mathrm{src}} \cdot \bm{v}_{\gas},
\end{aligned}
\end{equation}

\begin{equation}
\label{eq:CE_dust}
    \frac{\partial \rho_{\parti, ni}}{\partial t}+\nabla \cdot\left(\rho_{\parti, ni} \bm{v}_{\parti, n}+\mathscr{F}_{\mathrm{p, dif}, ni}\right)=0,
\end{equation}

\begin{equation}
\label{eq:ME_dust}
    \begin{aligned}
    \frac{\partial \rho_{\parti, ni}\left(\bm{v}_{\parti, n}+\bm{v}_{\parti, \mathrm{dif}, ni}\right)}{\partial t}+\nabla \cdot\left(\rho_{\parti, ni} \bm{v}_{\parti, n} \bm{v}_{\parti, n}+\bm{\Pi}_{\mathrm{dif}, ni}\right) & = \\
    \rho_{\parti, ni} \bm{f}_{\parti, \mathrm{src}, ni}+\rho_{\parti, ni} \frac{\bm{v}_{\mathrm{g}}-\bm{v}_{\parti, n}}{t_{\mathrm{s}, n}},
    \end{aligned}
\end{equation}

\begin{equation}
\label{eq:CE_vapor}
\frac{\partial \rho_{\tracer,i}}{\partial t}+\nabla \cdot\left(\rho_{\tracer, i} \bm{v}_{\gas}+\mathscr{F}_{\mathrm{tr, dif}, i}\right)=0.
\end{equation}
In these equations the index $n$ refers to the particle species and $i$ to the material component.
The particle concentration diffusion flux is given by,
\begin{equation}
    \label{eq:diffusion}
    \mathscr{F}_{\operatorname{p,dif}, ni} \equiv-\rho_{\mathrm{g}} D_{\parti, n} \nabla\left(\frac{\rho_{\parti, ni}}{\rho_{\mathrm{g}}}\right)
    \equiv \rho_{\parti, ni} \bm{v}_{\parti, \mathrm{dif}, ni}
\end{equation}
which defines the effective diffusive velocity $\bm{v}_{\textrm{p,diff},ni}$.  The particle diffusivity $D_{\parti,n}$ is a free parameter in our simulation related to the gas diffusivity $D_{\gas}$:
\begin{equation}
    \label{eq:D_particle}
    D_{\parti,n} = \frac{D_{\gas}}{1+\mathrm{St}_{n}^{2}}
\end{equation}
where $\mathrm{St}_n = t_{\mathrm{s},n}\Omega_{\mathrm{K}}$ is the dimensionless stopping time, i.e, the Stokes number \citep{YoudinLithwick2007}.
$\bm{\Pi}_{\mathrm{dif}, ni}$ is the momentum diffusion tensor, which describes the momentum flux caused by diffusion. Combined with $\bm{v}_{\textrm{p,diff},ni}$, \eq{ME_dust} ensures the Galilean invariance \citep{HuangBai2022}. And the last term of \eq{ME_dust} denotes the aerodynamic drag felt by particles, while we omit the backreaction on gas (see \citealt{HuangBai2022} for a complete form of \eqs{ME_gas}{EE} including dust feedback). \eq{CE_vapor} is the continuity equation of the vapour tracers whose velocity is the same as gas. Similar to \eq{diffusion}, the tracer concentration diffusion flux is defined as,
\begin{equation}
    \label{eq:D_tracer}
    \mathscr{F}_{\operatorname{tr, dif}, i} \equiv-\rho_{\gas} D_{\tracer} \nabla\left(\frac{\rho_{\tracer, i}}{\rho_{\mathrm{g}}}\right)
\end{equation}
where $D_{\tracer} \equiv D_{\gas}$ is the tracer diffusivity.

In \eqs{ME_gas}{ME_dust}  $\bm{f}_{\mathrm{src}}$ denotes the source term from external forces. Planet gravity is one of the external forces:
\begin{equation}
    \bm{f}_{\mathrm{planet}} = -\frac{GM_{\rm p}}{r^{2}}.
\end{equation}
Following \citet{MoldenhauerEtal2021}, we omit smoothing of the planet gravity since we aim to reveal flow patterns deep in the atmosphere. Also, we omit self-gravity. As our simulations are conducted in a local frame centred on the planet, two additional forces should be included in this non-inertial frame. The first is the Coriolis forces,
\begin{equation}
    \bm{f}_{\mathrm{cor}} = -2\bm{\Omega}_{K} \times \bm{v}.
\end{equation}
The second is the tidal force induced by the summation of stellar gravity and centrifugal force. Since we focus on a small domain near the planet, the tidal force can be expanded and approximated to the first order as,
\begin{equation}
    \bm{f}_{\mathrm{tid}} = (-GM_{\star}/a_{\mathrm{semi}}^{2} + \Omega_{\mathrm{K}}^{2} a_{\mathrm{semi}})\cdot \bm{e}_{x}  \approx 3\Omega_{\mathrm{K}}^{2} x \cdot \bm{e}_{x}
\end{equation}
where $\bm{e}_{x}$ denotes the direction pointing from the star to the planet and $x$ is the corresponding coordinate in this comoving frame. Here, $a_{\mathrm{semi}}$ is the semi-major axis of the planet (in circular orbit). Finally we have,
\begin{equation}
    \bm{f}_{\mathrm{src}} = \bm{f}_{\mathrm{cor}} + \bm{f}_{\mathrm{tid}} + \bm{f}_{\mathrm{planet}}.
\end{equation}

\subsection{Phase change model}
\label{sec:phase change model}
We adopt the ideal equation of state (EOS) for a multi-component fluid:
\begin{equation}
\label{eq:thermal-EOS}
    P = \sum_{i}^{N_{\gas}}\rho_{\gas, i} \frac{k_{\mathrm{B}}T}{\mu_{i} m_{\rm p}}= \rho_{\gas} \frac{k_{\mathrm{B}}T}{\mu_{\gas} m_{\rm p}}
\end{equation}
where $\rho_{\gas, i}$ is the density of gas component $i$, $\mu_{i}$ is the mean molecular weight of gas component $i$, $\rho_{\gas}$ is total density and $\mu_{\gas}$ is the mean molecular weight of the gas:
\begin{equation}
    \frac{1}{\mu_{\gas}} = \sum_{i}^{N_{\gas}}\frac{f_{\gas,i}}{\mu_{i}}
\end{equation}
where $f_{\gas,i} = \rho_{\gas, i} / \rho_{\gas} $ are the mass fraction of the gas phase components. These include the vaporized ices as well as the non-condensable materials and we denoted with $N_{\gas}$ the total number of them. In our case, we combine H and He into one non-condensable component so that $N_{\gas} = N_{\mathrm{z}}+1$.

To link the EOS to the specific energy of the gas, one needs to supply another caloric EOS, which is written as,
\begin{equation}
    \rho_{\gas} e_{\gas} = \frac{P}{\gamma - 1}
\end{equation}
where for an ideal fluid $\gamma$ is the heat capacity ratio and $e$ is the specific gas internal energy. For an ideal gas, the internal energy can be written as a summation of the specific heat of the different gas components plus the chemical potential of the vaporized ices,
\begin{equation}
\label{eq:rhoe_gas}
    \rho_{g} e_{\gas} = \sum_{i}^{N_{\gas}}C_{V\gas,i} \rho_{\gas,i} T + \sum_{i}^{N_{\mathrm{z}}} \rho_{z,i} \phi_{z,i}
\end{equation}
Similarly, for particles we have,
\begin{equation}
    \label{eq:rhoe_dust}
    \rho_{\parti} e_{\parti} =\sum_{n}^{N_{\parti}} \sum_{i}^{N_{\mathrm{z}}}C_{\parti,i}\rho_{\parti,ni}T + \sum_{n}^{N_{\parti}} \sum_{i}^{N_{\mathrm{z}}} \rho_{\parti,ni} \phi_{\parti,i}
\end{equation}
Here $C_{V\gas,i}$ is ideal gas heat capacity at constant volume and $C_{\parti,i}$ is the heat capacity of solids.  In practice, only the difference in chemical potential matters. It is convenient to set the chemical potential of vapour to be zero, $\phi_{z,i}=0$, such that the caloric EOS stays ideal. Also, we omit the chemical potential for non-condensable components. Then,
\begin{equation}
    \label{eq:gamma_general}
    \frac{1}{\gamma -1} = \mu_{\gas}\sum_{i}^{N_{\gas}} f_{\gas,i}\frac{C_{V\gas,i}}{k_{\mathrm{B}}/m_{\rm p}}
\end{equation}
where we combined \eqto{thermal-EOS}{rhoe_gas}.  Given the gas mass fractions $f_{\gas,i}$ and the heat capacity values, we hence obtain $\gamma$. Furthermore, under the ideal gas assumption, $\gamma$ equals the adiabatic index, which is needed in the Riemann solver.

The next step is to determine the fraction of each vapour component. During accretion, the planet atmosphere becomes much hotter than the ambient disc. For simplicity, we assume that phase change processes happen instantaneously and that all vapour components follow their saturation pressure in equilibrium. This assumption follows previous studies \citep{BrouwersOrmel2020,OrmelEtal2021}, but the sublimation process can alternatively be described by rate equations \citep{RosJohansen2013,SchoonenbergOrmel2017}. According to the \textit{Clausius-Clapeyron} equation, the saturation vapour pressure of species $i$ reads,
\begin{equation}
    \label{eq:saturated}
    P_{\mathrm{eq},i} = P_{\mathrm{eq0},i} \exp{(-T_{\mathrm{a},i}/T)}
\end{equation}
where $P_{\mathrm{eq0},i}$ and $T_{\mathrm{a},i}$ are constants specific to different materials.

To consistently deal with mass and energy transfer during phase change, we use the energy and mass conservation relationship in a local grid cell following \citet{LiChen2019}.  For a gas parcel containing various material components and particles, the total energy can be written as:
\begin{equation}
    \rho E = \rho_{\gas} e_{\gas} + \rho_{\gas} e_{K\gas} + \rho_{\parti} e_{\parti} + \sum_{n}^{N_{\parti}}\sum_{i}^{N_{\mathrm{z}}} \rho_{\parti,ni} e_{K\parti,n}
\end{equation}
Here $e_{K} = \frac{1}{2} \bm{v}^{2}$ denotes the specific kinetic energy. Note that all gaseous species have the same specific kinetic energy $e_{K\gas}$.  Moreover, latent heat is defined as the enthalpy difference between the vapour and the condensate:
\begin{equation}
    L_{\mathrm{heat},i}(T) = C_{P\gas,i}T + \phi_{z,i} - (C_{\parti,i}T + \phi_{\parti,i})
\end{equation}
where $C_{P\gas,i}$ is heat capacity at constant pressure and $L_{\mathrm{heat},i}$ is the specific latent heat for vapour species $i$. 
Therefore, the difference of the chemical potential is,
\begin{equation}
\label{eq:nu_difference}
    \Delta \phi_{i} = \phi_{z,i} - \phi_{\parti,i} =  L_{\mathrm{heat},i}(T) - \Delta C_{P,i} T
\end{equation}
where $\Delta C_{P,i} = C_{P\gas,i} -C_{\parti,i} $. Similarly, we can also define the specific kinetic energy difference between the solid and the gaseous phase $\Delta e_{K,n} = \left(e_{K\gas,n} - e_{K\parti}\right)$.

Let the mass transfer of material component $i$ from the solid phase to the gas phase be denoted $\delta \rho_{i}$ ($\delta\rho_{i}>0$ for sublimation; also, quantities involving $\delta$ are unknown and solved for by the vapour module, while $\Delta$-labelled quantities are constants or follow from the hydro step.). Furthermore, let the corresponding temperature change be $\delta T$. We need to partition $\delta\rho_{i}$ among the different particle species. For simplicity, we assume that $\delta \rho_{i}$ is distributed proportional to the mass fractions of the different particle species among material component $i$. That is,
\begin{equation}
    \label{eq:partition}
    \delta \rho_{\parti,ni} = -\delta \rho_{i} \frac{\rho_{\parti,ni}}{\sum_{n}^{N_{\parti}}\rho_{\parti,ni}}
\end{equation}
Then energy conservation yields,
\begin{equation}
\label{eq:DeltaT}
\begin{aligned}
    & \left(\sum_{i}^{N_{\gas}}C_{V\gas,i} \rho_{\gas,i}   + \sum_{n}^{N_{\parti}} \sum_{i}^{N_{\mathrm{z}}}C_{\parti,i}\rho_{\parti,ni} \right)T  = \\
    & \sum_{i}^{N_{\mathrm{z}}} \Delta \phi_{i}\delta \rho_{i} + \sum_{n}^{N_{\parti}} \sum_{i}^{N_{\mathrm{z}}} \Delta e_{K,n} \delta \rho_{\parti,ni} +  \left(T+\delta T\right) \times \\
    & \left[\sum_{i}^{N_{\gas}}C_{V\gas,i} \rho_{\gas,i}+ \sum_{n}^{N_{\parti}}\sum_{i}^{N_{\mathrm{z}}}C_{\parti,i}\rho_{\parti,ni} + \sum_{i}^{N_{\mathrm{z}}} \delta \rho_{i} \left( C_{V\gas,i} - C_{\parti,j} \right) \right] 
\end{aligned}
\end{equation}
In this equation, the LHS is the total gas and particle energy before sublimation. These terms also appear on the RHS. The additional terms on the RHS represent the energy exchange due to sublimation. The first term on the RHS is caused by the chemical potential difference, which mainly accounts for latent heat absorption ($\Delta \phi_{i}$ is usually positive). The second term on the RHS accounts for kinetic energy exchange due to the mass transfer. The last term on the RHS is due to both the temperature change and the heat capacity difference of the material components between gas and ice states.

\Eq{DeltaT} gives a relation between $\delta T$ and $\delta\rho_{i}$. In addition, each $\delta \rho_{i}$ should fulfill the material supply limit. That is, the changes are limited to:
\begin{equation}
\label{eq:supply}
    \delta\rho_i \ge -\rho_{z,i}\qquad \textrm{and} 
    \qquad 
    \delta \rho_{\parti,ni} \ge -\rho_{\parti,ni}.
\end{equation}
The first expression stipulates that we cannot freeze out more than the amount of available vapour, the second demands that we cannot sublimate all the available ice mass.

We can also include momentum conservation by assuming that sublimation happens isotropically so that the velocity of pebbles does not change. Therefore, the specific momentum of pebbles is conserved during phase change while the released vapour alters the momentum of the gas parcel during its mixing. This process can be described as,
\begin{equation}
    \label{eq:mom_conserve}
    \begin{aligned}
    &(\rho_{\gas} + \sum_{i}^{N_{\mathrm{z}}} \delta \rho_{i})(\bm{v}_{\gas} + \delta \bm{v}_{\gas}) -\rho_{\gas} \bm{v}_{\gas} +  \sum_{n}^{N_{\parti}} \sum_{i}^{N_{\mathrm{z}}} \delta \rho_{\parti, ni} \bm{v}_{\parti, n} = 0 
    \end{aligned}
\end{equation}
where $\delta \bm{v}_{\gas}$ is the corresponding gas velocity change during the phase change process. This equation can be solved for $\delta\bm{v}_\mathrm{g}$ and the gas specific kinetic energy $e_{K\gas}$ is accordingly modified. Since in our simulation setup we only consider vapour freeze-out on tiny, well-coupled ice grains, where $\bm{v}_{\parti,n} = \bm{v}_{\gas}$ and $\delta \bm{v}_{\gas} = 0$, momentum conservation is guaranteed. In general, however, e.g., with vapour freezing out on pebbles that drift with respect to the gas, an equation similar to \eq{mom_conserve} applicable to the condensation process, needs to be invoked.

In \eq{DeltaT}, there are $(1+i)$ unknowns ($\delta T$ and $\delta \rho_{i}$). However, recall that we also have $i$ vapour saturation pressure equations for all ice components (\eq{saturated}). Thus in total there are $(1+i)$ unknowns and $(1+i)$ equations. We numerically solve for the unknowns using the standard secant method \citep{PressEtal1986}. In this procedure, the root is initialized with the values from the hydrodynamical step. By varying $\delta T$, $\delta \rho_{i}$ is acquired from the saturation curve and the root is found when the energy conservation is met simultaneously (\eq{DeltaT}). 

\subsection{Implementation in hydrodyanmical codes}
\label{sec:implement}
Unless specified otherwise, our default setup contains one non-condensable H-He mixture, one water ice and one refractory (silicate). Their material properties are listed in \tb{sublimation_const}. In this setup, we need one gas fluid, one tracer fluid and two dust fluid.

\subsubsection{PLUTO}
\label{sec:pluto_implement}
The public version of the PLUTO code \citep{MignoneEtal2007,MignoneEtal2012} does not accommodate a dust-fluid module so that we cannot follow the hydrodynamics of pebbles directly. However, it contains a built-in chemical module. In this module the material component fractions are specified and can be updated for each timestep. For example, H-He and water vapour are two chemical components with different mean molecular weight. Thus for well-coupled ice grains ($\mathrm{St} \longrightarrow 0$) which exactly follow the gas, we can also treat them as another chemical components with their mean molecular weight infinity. With this setup, PLUTO is capable to simulate special cases with well-coupled ice grains. Due to this limitation, we mainly run PLUTO to compare with the results from Athena++ and verify the design of our phase change model.

\subsubsection{ATHENA++}
To fully capture the dynamics of pebbles as well as hydrodynamic behavior of gas simultaneously, we employ the recently developed multi-fluid dust module in Athena++ \citep{HuangBai2022}. In this module, dust is modelled as a pressureless fluid. An arbitrary number of dust species, interacting with the gas via aerodynamic drag regulated by their stopping time, can be specified. Apart from the dust fluids, this module accommodates only one single gas fluid. Thus, to follow the vapour, we add several tracer fluids (see \se{design} for details). In addition, we have modified the module to allow mass transfer among the components, in accordance to the phase change model \se{phase change model}.

The hydrodynamic part is calculated by the multi-fluid dust module, which solves \eqto{CE_gas}{ME_dust}. The thermodynamic part is calculated according to the phase change model described in \se{phase change model}.
Since the thermodynamic calculations also update the local sound speed, the timestep is affected by the phase change model via the Courant--Friedrichs--Lewy (CFL) condition for numerical stability.

\subsubsection{Specification of thermodynamic quantities}
In principle, $C_{P\gas,i}$, $C_{V\gas,i}$, $C_{\parti,i}$ and $L_{\mathrm{heat},i}$ all depend on temperature. Their precise value can be obtained from experiments combining with fittings \citep{FraySchmitt2009}, where higher order temperature terms are also extended based on the classical \textit{Clausius-Clapeyron} equation. Hydrogen and helium as simple molecules have their heat capacity well approximated with the ideal gas:
\begin{equation}
    \label{eq: Cv}
    C_{V\gas,i} = \frac{n_{i}}{2} \frac{k_{\mathrm{B}}}{\mu_{i} m_{\rm p}}
\end{equation}
where $n_{i}$ is the degree of freedom of motion. On the other hand, water deviates more from this heat capacity relation \citep{FraySchmitt2009}. However, in this paper we focus on demonstrating hydrodynamic effect of phase change rather than digging into precise thermodynamic properties, thus we apply simple heat capacity relationship that writes as, 
\begin{equation}
\begin{aligned}
    C_{P\gas,i} &= \frac{k_{\mathrm{B}}}{\mu_{i} m_{\mathrm{p}}} + C_{V\gas,i} \\
    C_{\parti,i} &= 3 \frac{k_{\mathrm{B}}}{\mu_{i} m_{\mathrm{p}}}
\end{aligned}
\end{equation}
Here for solid heat capacity $C_{\parti, i}$, we have assigned six degrees of freedom of motion (three transverse and three rotational, $n_{i} =6$).

Then \eq{gamma_general} simplifies as,
\begin{equation}
    \label{eq:gamma_simplify}
    \frac{1}{\gamma -1} = \mu_{\gas}\sum_{i}^{N_{\gas}} \frac{f_{\gas,i}}{2}\frac{n_{i}}{\mu_{i}}
\end{equation}

For an ideal gas, the latent heat as a function of temperature can be expressed as:
\begin{equation}
\begin{aligned}
    L_{\mathrm{heat},i}(T) &= \int_{T_r}^{T} - \Delta C_{P,i}(T) ~ dT +  L_{\mathrm{heat},i}(T_{r}) \\
    & \approx   L_{\mathrm{heat},i}(T_{r}) - \Delta C_{P,i}(T)(T-T_{r})
\end{aligned}
\end{equation}
where $T_{r}$ is the reference temperature. In the temperature range we are interested in, the experimental results and analytic solutions show that $L_{\mathrm{heat},i}(T)$ varies little but has large systematic uncertainties \citep{Feistel2007SublimationPA}.
Considering this and our focusing on hydrodynamics, we decide to use a constant latent heat value in simulation. Thus \eq{nu_difference} reduces to:
\begin{equation}
    \Delta \phi_{i} = \phi_{\parti,i} - \phi_{z,i} = - L_{\mathrm{heat},i}(T_{r}) 
\end{equation}
Here after we omit $T_{r}$ and list the value used in our simulation with $L_{\mathrm{heat}}$ in \tb{sublimation_const}.
\subsection{disc model}
\label{sec:disc_model}
Throughout the simulation runs, we adopt the Minimum Mass Solar Nebula (MMSN) disc model \citep{Weidenschilling1977,Hayashi1981}. The surface density and temperature profile read:
\begin{equation}
    \Sigma(a_{\mathrm{semi}}) = 1.7 \times 10^{3} \left(\frac{a_{\mathrm{semi}}}{1 ~\si{au}}\right)^{-3/2} \si{g~cm^{-2}}
\end{equation}
\begin{equation}
    \label{eq:disc_T}
    T(a_{\mathrm{semi}}) = 270 \left(\frac{a_{\mathrm{semi}}}{1 ~\si{au}}\right)^{-1/2} \si{K}
\end{equation}


The mass of the planetary core is characterized by a dimensionless parameter $m = M_{\mathrm{p}}/M_{\mathrm{th}}$, where $M_{\mathrm{th}} = c_{\rm s,iso}^{3}/(G \Omega_{\mathrm{K}})$ is the thermal mass. Here $c_{\rm s,iso}$ is the isothermal sound speed corresponding to the disc temperature. In the MMSN disc, the dimensionless mass $m$ and the planetary core mass are related as follows,
\begin{equation}
    \frac{M_{\mathrm{p}}}{M_{\oplus}} \approx 11.8~ m ~\left(\frac{a_{\mathrm{semi}}}{1~\text{au}}\right)^{3/4}.
\end{equation}
Also, the Bondi radius of the planet is defined as,
\begin{equation}
    \label{eq:bondi_radius}
    R_{\text{Bondi}} = \frac{G M_{\mathrm{p}}}{c_{\rm s,iso}^{2}} \equiv m H_{\gas}
\end{equation}
where $H_{\gas}= c_{\rm s,iso} /\Omega$ is the local scale height at the planet's orbital radius.
Unless further specified, we assume an inviscid gas flow following previous studies \citep{OrmelEtal2015,BethuneRafikov2019,KurokawaTanigawa2018}. However, in order to speed-up the simulation convergence we do adopt a small but non-zero diffusivity for the vapour and ice grains.
\begin{equation}
    D_{\gas} = \beta c_{\rm s,iso} H_{\gas}
\end{equation}
Here we take $\beta = 10^{-5}$ for our simulation (see \tb{simu_paras}).

Unless specified, we use a set of dimensionless units to show the results, where the length is in unit of local scale height $H_{\gas}$, time is in unit of $\Omega_{\mathrm{k}}^{-1}$ and density is in units of the disc midplane density at planet position: $\rho_{\gas,0} = \Sigma/(\sqrt{2\pi} H_{\gas})$. The temperature is still expressed in Kelvin.

\begin{table*}
	\centering
    \caption{Parameters of all simulations runs. Columns denote $a_{\mathrm{semi}}$:semi-major axis of planets; $m$: dimensionless planet mass (see \se{disc_model}); $N_{\mathrm{\parti}}$: number of particle species; material: material components included in pebbles; $f_{z}$: initial material component fractions for the particle species with non-zero St; St: dimensionless stopping time of each particle species with respect to the background disc; $f_{\mathrm{p2g,ini}}$: initial pebble-to-gas ratio; $D_{\gas}$: diffusivity of gas and vapour in terms of $H^2\Omega^{-1}$; $t_{\mathrm{soft}}$: planet gravity softening time; RES: resolution of simulation ($N_{r}\times N_{\phi}$); domain: simulation domain $(r_{\mathrm{min}}, r_{\mathrm{max}}) \times (\phi_{\mathrm{min}}, \phi_{\mathrm{max}})$ in unit of local gas scale height ($H_{\gas}$) and radian.}
	\label{tab:simu_paras}
	\begin{tabular}{lccccccccccccc} 
		\hline
		simulation name &  $a_{\mathrm{semi}}$ & m & $N_{\mathrm{\parti}}$  & material & $f_{z}$ & St & $f_{\mathrm{p2g,ini}}$ & ${D}_{\mathrm{\gas}}$ & $t_{\mathrm{soft}}$ & RES & domain\\

          & (au) & & & & & & & ($H^2\Omega^{-1}$) & $(\Omega_{\mathrm{K}}^{-1})$ & & ($H_{\gas} \times \si{rad}$) \\
		\hline
		\texttt{gasOnly} & 5.0 & 0.1 &\textbackslash & \textbackslash & \textbackslash & \textbackslash & \textbackslash & \textbackslash & 2 & 256$\times$256 & ($10^{-3}$,1.0)$\times$(0,$2\pi$) \\
		\texttt{coupled} & 5.0 & 0.1 & 1 & (b) & 1.0 & 0 & 0.25 & 0.0 & 2 & 256$\times$256 & ($10^{-3}$,1.0)$\times$(0,$2\pi$) \\
		\texttt{coupled-Pluto} & 5.0 & 0.1  & 1 & (b) & 1.0 & 0 & 0.25 & 0.0 & 2 & 256$\times$256 & ($10^{-3}$,1.0)$\times$(0,$2\pi$)\\
		\texttt{pebble-fiducial} & 4.0 & 0.1 & 2 & (b),(c) & 0.5;0.5 & 0.01; 0 & 0.02 & $10^{-5}$ & 5 & 128$\times$64 & ($5 \times 10^{-3}$,1.0)$\times$(0,$\pi$)\\
		\texttt{pebble-3.5au} & 3.5 & 0.1 & 2 & (b),(c) & 0.5;0.5 & 0.01; 0 & 0.02 & $10^{-5}$ & 5 & 128$\times$64 & ($5 \times 10^{-3}$,1.0)$\times$(0,$\pi$) \\
        \texttt{pebble-4.5au} & 4.5 & 0.1 & 2 & (b),(c) & 0.5;0.5 & 0.01; 0 & 0.02 & $10^{-5}$ & 5 & 128$\times$64 & ($5 \times 10^{-3}$,1.0)$\times$(0,$\pi$) \\
		\texttt{pebble-5au} & 5.0 & 0.1 & 2 & (b),(c) & 0.5;0.5 & 0.01; 0 & 0.02 &  $10^{-5}$ & 5 & 128$\times$64 & ($5 \times 10^{-3}$,1.0)$\times$(0,$\pi$) \\
        \texttt{pebble-6au} & 6.0 & 0.1 & 2 & (b),(c) & 0.5;0.5 & 0.01; 0 & 0.02 &  $10^{-5}$ & 5 & 128$\times$64 & ($5 \times 10^{-3}$,1.0)$\times$(0,$\pi$) \\
        \texttt{pebble-p2g0.01} & 4.0 & 0.1 & 2 & (b),(c) & 0.5;0.5 & 0.01; 0 & 0.01 & $10^{-5}$ & 5 & 128$\times$64 & ($5 \times 10^{-3}$,1.0)$\times$(0,$\pi$) \\
        \texttt{pebble-p2g0.05} & 4.0 & 0.1 & 2 & (b),(c) & 0.5;0.5 & 0.01; 0 & 0.05 & $10^{-5}$ & 5 & 128$\times$64 & ($5 \times 10^{-3}$,1.0)$\times$(0,$\pi$) \\
        \texttt{pebble-p2g0.1} & 4.0 & 0.1 & 2 & (b),(c) & 0.5;0.5 & 0.01; 0 & 0.1 & $10^{-5}$ & 5 & 128$\times$64 & ($5 \times 10^{-3}$,1.0)$\times$(0,$\pi$) \\
        \texttt{pebble-St0.03} & 4.0 & 0.1 & 2 & (b),(c) & 0.5;0.5 & 0.03; 0 & 0.02 & $10^{-5}$ & 5 & 128$\times$64 & ($5 \times 10^{-3}$,1.0)$\times$(0,$\pi$) \\
        \texttt{pebble-St0.003} & 4.0 & 0.1 & 2 & (b),(c) & 0.5;0.5 & 0.003; 0 & 0.02 & $10^{-5}$ & 5 & 128$\times$64 & ($5 \times 10^{-3}$,1.0)$\times$(0,$\pi$) \\
        \texttt{pebble-m0.05} & 4.0 & 0.05 & 2 & (b),(c) & 0.5;0.5 & 0.01; 0 & 0.02 & $10^{-5}$ & 5 & 128$\times$64 & ($5 \times 10^{-3}$,1.0)$\times$(0,$\pi$) \\
        \texttt{pebble-m0.2} & 4.0 & 0.2 & 2 & (b),(c) & 0.5;0.5 & 0.01; 0 & 0.02 & $10^{-5}$ & 5 & 256$\times$128 & ($5 \times 10^{-3}$,1.0)$\times$(0,$\pi$) \\
        \texttt{pebble-m0.2-4.5au} & 4.5 & 0.2 & 2 & (b),(c) & 0.5;0.5 & 0.01; 0 & 0.02 & $10^{-5}$ & 5 & 256$\times$128 & ($5 \times 10^{-3}$,1.0)$\times$(0,$\pi$) \\
		\hline
	\end{tabular}
\end{table*}
\subsection{Pebble dynamics}
We include two drag regimes: the Epstein regime and the Stokes regime drag. These drag regimes are determined by the size of the pebble in relation to the mean free path of the gas molecules $l_{\mathrm{mfp}}$,
\begin{equation}
    l_{\mathrm{mfp}}=\frac{\mu_{\gas} m_{\rm p}}{\sqrt{2} \rho_{\mathrm{g}} \sigma_{\mathrm{mol}}}.
\end{equation}
Here, $\sigma_{\mathrm{mol}}$ is the molecular collision cross-section, which we approximate by the collision cross-section of molecular hydrogen $\sigma_{\mathrm{mol}}=2 \times 10^{-15} \mathrm{~cm}^{2}$ \citep{ChapmanCowling1991}. The stopping time is given by (we drop the subscript $n$),
\begin{equation}
    t_{\mathrm{s}}=\left\{
    \begin{aligned}
    \frac{\rho_{\bullet, \parti} s_{\parti}}{v_{\mathrm{th}} \rho_{\gas}} & , & (\mathrm{Epstein:}  \quad s_{\mathrm{p}}<\frac{9}{4} l_{\mathrm{mfp}}  ), \\
    \frac{4 \rho_{\bullet, \parti} s_{\parti}^{2}}{9 v_{\mathrm{th}} \rho_{\gas}l_{\mathrm{mfp}}} & , & (\mathrm{Stokes:}  \quad s_{\mathrm{p}}>\frac{9}{4} l_{\mathrm{mfp}}  ).
    \end{aligned}
    \right.
\end{equation}
where $s_{\parti}$ is the particle size, $\rho_{\bullet, \parti}$ is the particle internal density and $v_{\mathrm{th}} = \sqrt{8/\pi}c_{\mathrm{s}}$ is the thermal velocity of the gas molecules. Here $c_{\rm s}$ is the local sound speed. For a particle compound, its internal density $\rho_{\bullet, \parti}$ is given by,
\begin{equation}
    \rho_{\bullet, \parti} = \left( \sum_{i}^{N_{z}} \frac{f_{\parti,i}}{\rho_{\bullet, \parti,i}} \right)^{-1}
\end{equation}
where $f_{\parti,i} = \rho_{\parti,i}/(\sum_{i}^{N_{z}}\rho_{\parti,i})$ is the fraction of material component $i$ and $\rho_{\bullet, \parti,i}$ is the internal density of material component $i$ (see \tb{sublimation_const} for detailed value).

In order to calculate stopping time, we need to compute the particle size $s_{\parti}$. Since the mass of the refractory component of each particle species is unaffected by the phase change process, we obtain the particle number density $n_{\parti}$ accordingly,
\begin{equation}
    n_{\parti} =\frac{\rho_{\parti,\mathrm{refrac}}}{m_{\parti,\mathrm{refrac,0}}}
\end{equation}
where $m_{\parti,\mathrm{refrac,0}}$ is the mass of refractories in this particle species and $\rho_{\parti,\mathrm{refrac}}$ the corresponding internal density. From the number density, the particle mass is obtained, $m_{\parti} = \sum_i^{Nz} \rho_{\parti,i}/n_\parti $,  which, combined with its internal density, gives $s_{\parti}$:
\begin{equation}
    \frac{4\pi}{3} s_{\parti}^3 = \frac{1}{n_{\parti}} \frac{\sum_{i}^{N_{z}} \rho_{\parti,i}}{\rho_{\bullet, \parti}} 
\end{equation}

\subsection{Model validation}
To verify the hydrodyanmical part of our model, we reproduce the case that inviscid gas flow composed of pure H and He passes through a planetary atmosphere, a problem that has already been studied before \citep{OrmelEtal2015,FungEtal2015,BethuneRafikov2019}. By design, the gas is adiabatic and $\gamma = 1.43$ (assume $71\%$ $\mathrm{H_{2}}$ and $29\%$ He). We also run cases with $\gamma = 1.3, 1.1$. Overall the flow patterns show the typical horseshoe orbit and inner rotational atmosphere that are consistent with previous studies (\fg{gasonly_pattern}). We also reproduce the trend that the atmosphere expands with $\gamma$ decreasing. However, there emerges unphysical entropy violation possibly caused by numerical viscosity for the run $\gamma =1.1$ (\fg{gasonly_profile}). Please refer to \se{gas_only} for detailed results and discussions.

To test the robustness of the phase change module, we apply it to the 1D snowline model where \citet{SchoonenbergOrmel2017} investigated the feasibility of planetesimal formation by streaming instability near the water snowline. They found that the solid surface density would be enhanced through vapour diffusing outwards across the snowline to recondense back to ices. We have also successfully reproduced the ice density enhancement just outside the snowline seen in their fiducial model. 

\begin{table*}
    \centering
    \caption{Thermodynamic constants of different material components. The given ID refers to the ``material'' column in \tb{simu_paras}.}
    \label{tab:sublimation_const}
    \begin{threeparttable}
        \begin{tabular}{lcccccc}
        \hline
        material & ID & $T_{\mathrm{a}}$ & $P_{\mathrm{eq,0}}$ & $L_{\mathrm{heat}}$ & $\mu$ & $\rho_{\bullet, \parti,i} $ \\
        & & $(\si{K})$ & $(\si{g ~cm^{-1} s^{-2}})$ & $(10^{7}\si{erg ~g^{-1}})$ & & $(\si{g ~cm^{-2}})$  \\
        \hline
        \texttt{H-He}& (a) &  - & - & - & 2.34 & -  \\
        \texttt{water} \tnote{1} & (b) & $5996$ & $1.14 \times 10^{13}$ &  $\sim  2750$ & 18.0 & 1.0   \\
        \texttt{silicate} \tnote{2} & (c)   & $22519$ & $1.27\times 10^{8}$ & $\sim$ 4179 & - & 3.0  \\
        \texttt{Pyrene} \tnote{3} &(d) & $11760$ & $7.44\times 10^{14}$  & $\sim$ 483 &  - & -   \\
        \hline
        \end{tabular}
        
        \begin{tablenotes}[para,online]
            \footnotesize\smallskip%
            \item[1] \citet{FraySchmitt2009}
            \item[2] \citet{CostaEtal2017}
            \item[3] \citet{Goldfarb2008VaporPA}
        \end{tablenotes}
    \end{threeparttable}
\end{table*}
\begin{figure*}
    \includegraphics[width=\textwidth]{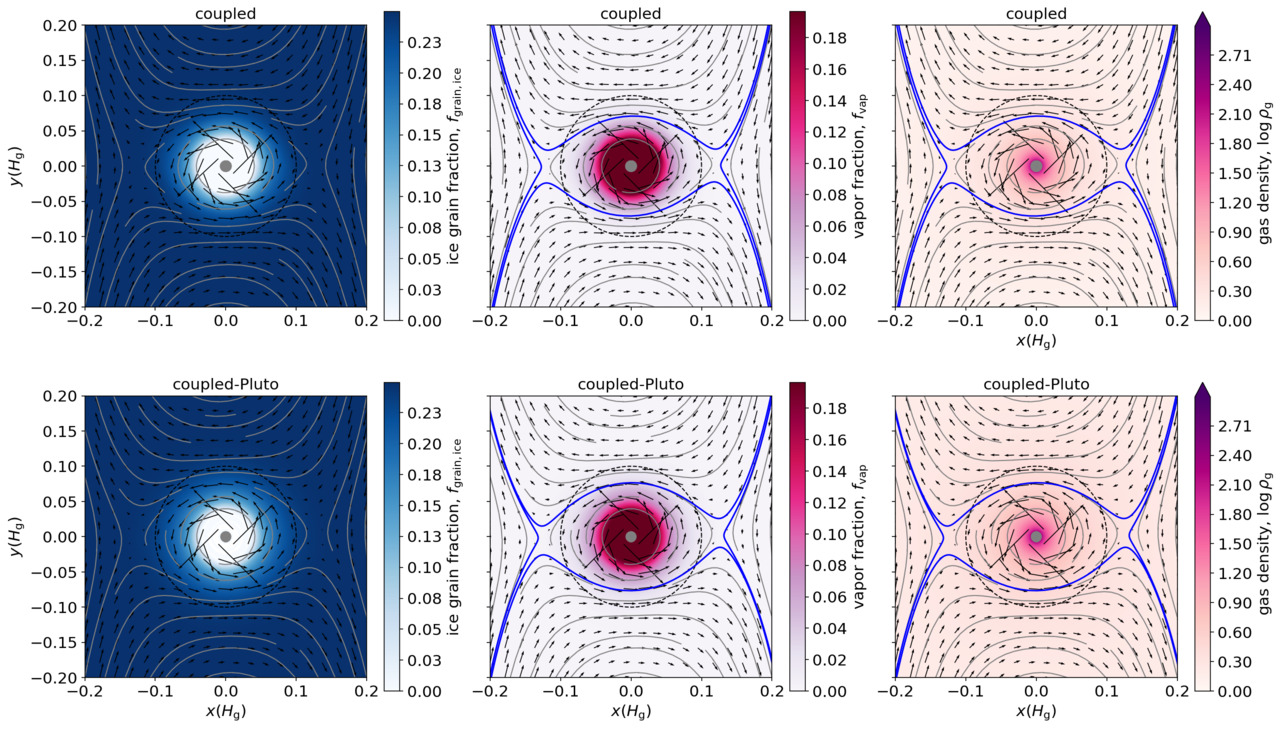}
    \caption{Comparison between well-coupled cases for the Athena++ (upper panels) and PLUTO (lower) runs (\texttt{coupled} and \texttt{coupled-Pluto}). The color in three columns denotes ice grain fraction ($f_{\mathrm{grain,ice}}$), vapour fraction ($f_{\mathrm{vap}}$) and gas density in logarithmic scale, respectively. The text in the title indicates the simulation name (\tb{simu_paras}). The critical streamline for gas (blue) is plotted in the middle and right columns (\se{coupled}).}
    \label{fig:coupled_contra}
\end{figure*}

\section{Results}
\label{sec:results}
In this section, we conduct simulations with either well-coupled grains or pebbles of non-zero Stokes number. All runs are listed in \tb{simu_paras}. For non-zero Stokes number runs, we name them with the prefix ``\texttt{pebble}'' and the text following the dash represents the exclusive parameters that are varied in this run. For example, \texttt{pebble-fiducial} represents the fiducial run where the semi-major axis of the planet ($a_{\mathrm{semi}}$) is 4.0 au. Then \texttt{pebble-3.5au} means the planet is put at 3.5 au while other parameters are kept the same as the fiducial run. The number of particle species included in this run is given by $N_{\parti}$. For each particle species we list their Stokes numbers ($\mathrm{St}$) and initial pebble-to-gas mass ratio ($f_{\mathrm{p2g,ini}}$). The Material column represents the material components contained in each particle species and their properties and corresponding ID are listed in \tb{sublimation_const}. Specifically, (b) represents water and ETC. For each material component, the initial mass fraction is given by $f_{z}$. For example, for the fiducial run, only a single particle species with $f_{\mathrm{p2g,ini}} = 0.02$ is initialized in the domain and water ice accounts for $50\%$ of it.

\subsection{Simulation domain and setup}
\label{sec:simu_set_up}
To better capture the hydrodynamics near the core, a polar coordinate with a logarithmically spaced grid in the radial direction is adopted \citep{OrmelEtal2015}, while in the azimuthal direction the grid is evenly spaced.  With this design the outer boundary can be placed far enough to not feel the perturbations by the planet. Therefore, we use a fixed outer boundary with constant density and a Keplerian shear velocity (both for pebbles and gas). As a consequence, gas and pebbles are replenished continuously from the outer boundary. For the inner boundary, if not specified otherwise, we use reflective boundary condition for gas and outflow for pebbles. We list our simulation resolution for all simulation runs in Table~\ref{tab:simu_paras}.

\begin{table*}
	\centering
    \caption{Summary of several characteristic outputs of all runs. Columns denote: $R_{\mathrm{subl}}$: radius of the sublimation front; $R_{\mathrm{atm}}$: atmosphere radius; $f_\mathrm{vap,peak}$: peak vapour fraction; $M_{\mathrm{vap}}$: total vapour mass in the whole domain;  $M_{\mathrm{vap,bound}}$: bound vapour mass inside $R_{\mathrm{atm}}$.  In status, we summary the runs's behavior by three regimes defined in the main text (recycling-dominated as ``R'', vapour-dominated as ``V'' and the intermediate regime as ``I''). 
    For the runs that have already reached a steady state, we list their steady value of all output quantities (take the average of last 10 $\Omega^{-1}$) while for the other runs (either quasi-steady or not steady), we list their value for the final simulation time.}
	\label{tab:outputs}
	\begin{tabular}{lcccccccc} 
		\hline
		simulation name &  $R_{\mathrm{subl}} (H_{\gas})$ & $R_{\mathrm{atm}} (H_{\gas})$ & $f_{\mathrm{vap,peak}}$  & $M_{\mathrm{vap}}(M_{\oplus})$ &  $M_{\mathrm{vap,bound}}(M_{\oplus})$ & status & simulation time($\Omega^{-1}$) \\
		\hline
		\texttt{pebble-fiducial} & 0.12 & 0.077 & 0.012 & 7.1$\times 10^{-4}$ & 2.4$\times 10^{-4}$ & R & 150\\
		\texttt{pebble-3.5au} & 0.20 & 0.079 & 0.010 & 0.0016 & 2.0$\times 10^{-4}$ & R  & 150\\
        \texttt{pebble-4.5au} & 0.052 & 0.081 & 0.045 & 0.0013 & 0.0010 & I  & 300\\
		\texttt{pebble-5au} & 0.018 & 0.13 & 0.37 & 0.027 & 0.027 & V  & 900\\
        \texttt{pebble-6au} & 0.018 & 0.13 & 0.34 & 0.022 & 0.022 & V  & 300\\
        \texttt{pebble-p2g0.01} & 0.15 & 0.079 & 0.0054 & 4.2$\times 10^{-4}$ & 1.1$\times 10^{-4}$  & R & 150\\
        \texttt{pebble-p2g0.05} & 0.054 & 0.086 & 0.073 & 0.0026 & 0.0020& I & 300\\
        \texttt{pebble-p2g0.1} & 0.021 & 0.13 & 0.43 & 0.039 & 0.037 & V & 300\\
        \texttt{pebble-St0.003} & 0.13 & 0.080 & 0.010 & 6.3$\times 10^{-4}$ & 2.1$\times 10^{-4}$ & R & 160\\
        \texttt{pebble-St0.03} & 0.059 & 0.10 & 0.057 & 0.0022 & 0.0018 & I & 900\\
        \texttt{pebble-m0.05} & 0.024 & 0.059 & 0.094 & 9.0$\times 10^{-4}$ & 7.1$\times 10^{-4}$& I & 400\\
        \texttt{pebble-m0.2} & 0.23 & 0.11 & 0.010 & 0.0017 & 5.8$\times 10^{-4}$ & R & 160\\
        \texttt{pebble-m0.2-4.5au} & 0.18 & 0.11 & 0.011 & 0.0015 & 7.5$\times 10^{-4}$ & R & 150\\
		\hline
	\end{tabular}
\end{table*}

\begin{figure}
    \includegraphics[width=\columnwidth]{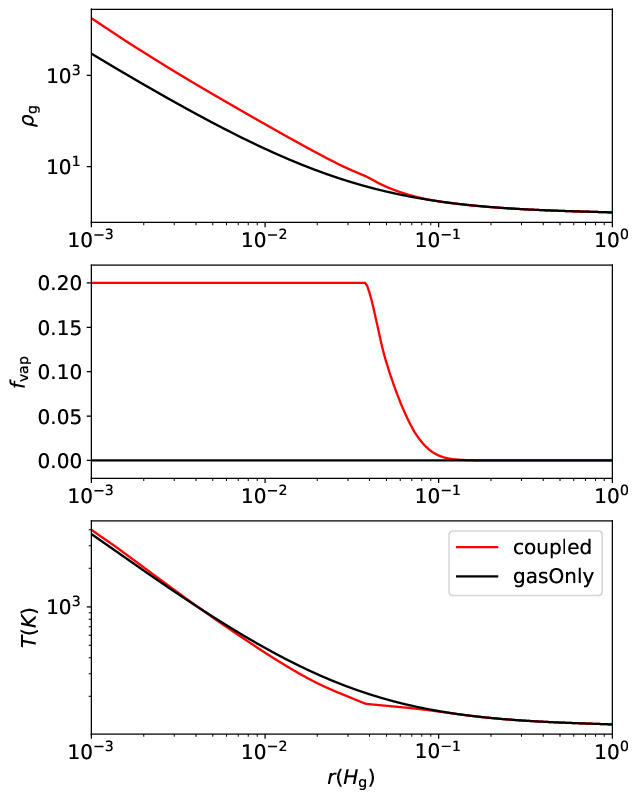}
    \caption{Comparison of azimuthally averaged quantities of runs \texttt{gasOnly} and \texttt{coupled}. From top to bottom, figures show profiles of midplane gas density, vapour fraction and temperature.}
    \label{fig:coupled_contra_profile}
\end{figure}

\begin{figure*}
    \includegraphics[width=\textwidth]{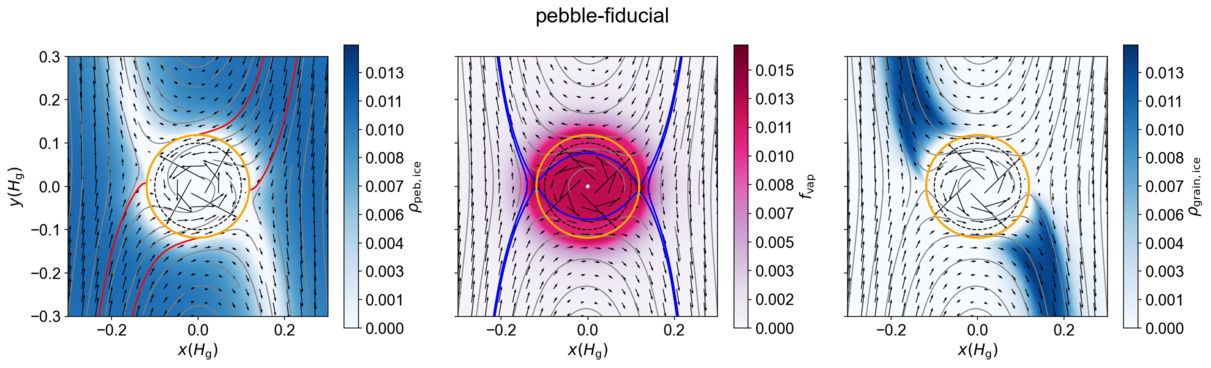}
    \caption{Flow patterns of run \texttt{pebble-fiducial}. From left to right, panels depict ice density contained in pebbles, the H$_2$O vapour fraction, and the ice density contained in grains. The black dashed line in each panel denotes the Bondi radius and the orange circle denotes the sublimation front (\se{fiducial}). The critical streamline (\se{coupled}) for gas (blue) is plotted in the middle panel (vapour) while the ``ice sublimation band'' (\se{fiducial}) where pebbles sublimate is denoted in the left panel by the red streamlines. A movie showing the temporal evolution is available in the online supplementary material and on github (\href{https://github.com/rainkings/Yu_pre/blob/main/videos/supplementary/pebble-fiducial.mp4}{pebble-fiducial}). }
    \label{fig:inject_fiducial}
\end{figure*}

In order to have a smooth start without shocks near the planet surface, a gravitational softening coefficient $f_{\mathrm{soft}}$ multiplying the planet's gravity is introduced. Here, we choose
\begin{equation}
    \label{eq:g_softening}
    f_{\mathrm{soft}} = 1 - \exp{[-0.5 (t/t_{\mathrm{soft}})^{2}]}
\end{equation}
following \citet{OrmelEtal2015,MoldenhauerEtal2021}, where $t_\textrm{soft}$ is the gravitational softening time.

In the $\mathrm{St} = 0$ simulations, we only consider one particle species: the well-coupled grains which contain only one material component, water ice. Thus in total the simulation contains one dust fluid. In addition, one tracer fluid is needed to follow the vapour density. Initially, well-coupled grains ($\mathrm{St} = 0$) are present in the domain at a pebble-to-gas ratio $f_\mathrm{p2g,ini}$ (\tb{simu_paras}). Also grains with same pebble-to-gas ratio $f_{\mathrm{p2g,ini}}$ continuously replenish from the outer boundary, rendering a steady state solution possible. Therefore, it is expected that the vapour fraction in the inner region (near the core) should level off at $(f_{\mathrm{p2g,ini}})/(1.0+f_{\mathrm{p2g,ini}})$. 

In the $\mathrm{St} > 0$ simulations, we consider two material components: (volatile) water ice and (refractory) silicates, along with the non-condensible H/He gas mixture (assume $71\%$ $\mathrm{H_{2}}$ and $29\%$ He). For simplicity, we assume that these materials are contained into pebbles of a single size (or Stokes number), of which only the H$_2$O ice can sublimate. We further assume that the water vapour will recondense onto grains that are essentially massless, $\textrm{St}=0$. Hence, there are two particle species -- pebbles and ice grains -- represented by in total three dust fluids. Also, one tracer fluid is added to follow the vapour density. For simplicity, we omit condensation of water vapour on pebbles and other processes as grain growth and fragmentation.

In contrast to the gas-only and perfectly-coupled simulations, we found that the $\textrm{St}>0$ simulations must be carefully initialized to avoid any spurious instabilities or vortices triggered by the presence of decoupled pebbles near the planetary core. Thus an ``relaxing'' boundary condition was introduced, which starts from an inflow/outflow boundary but gradually relaxes to reflective. Specifically, the velocity at the ghost cell is written as,
\begin{equation}
    v_{\gas,r}^{\prime} = -2v_{\gas,r} f_{\mathrm{soft}} + v_{\gas,r}
\end{equation}
where $v_{\gas,r}^{\prime}$ is the radial gas velocity of the ghost zone and $f_{\mathrm{soft}} = 1 - \exp{[-0.5 (t/t_{\mathrm{soft}})^{2}]}$, which is the same as used for the gravitational softening factor (\eq{g_softening}). 
In our runs, we choose $t_{\mathrm{soft}} = 5~\Omega_{\mathrm{K}}^{-1}$ and larger $t_{\mathrm{soft}}$ only bring slight difference on the steady state vapour fraction, which will not influence our conclusions.

In \app{gas_only} we report an entropy violation issue, possibly induced by the high azimuthal velocity . For this reason, we increase the size of the inner radius with respect to the \texttt{gasOnly} and \texttt{coupled} runs and lower the resolution accordingly in the $\mathrm{St}\neq0$ simulations (\tb{simu_paras}). Moreover, we conduct only half-domain simulations where only $\phi = [0,\pi]$ is simulated considering that all the forces and initial conditions are symmetric about the central planet. In runs that nevertheless yielded unacceptably high numerical viscosity, we have increased the resolution (for example, \texttt{pebble-m0.2}). 

\subsection{Strongly coupled dust runs}
\label{sec:coupled}
Since well-coupled grains follow the gas, we can implement them both in PLUTO with Chemical module (\se{pluto_implement}) and Athena++ with the multifluid dust module. By comparing their outputs, we are able to verify the robustness of the phase change module presented in \se{phase change model}.

In \fg{coupled_contra}, we show the results from both PLUTO and Athena++ after a steady state has emerged for the run with $f_{\mathrm{p2g,ini}} = 0.25$ (\tb{simu_paras}, \texttt{coupled} and \texttt{coupled-Pluto}). The flow pattern for the PLUTO and Athena++ runs are nearly identical. They both show symmetric horseshoe orbits and central rotational atmospheres. Near the core where the environment is hot, ice grains (left) sublimate and convert to vapour (centre). Further from the core, no vapour exists. There is a sharp increase in the vapour fraction as well as a sharp decrease in the ice grain fraction, which are due to the exponential form of the saturation profile (\eq{saturated}). Following the horseshoe orbit, some vapour outside the blue streamline can be transported to cooler region and freeze out immediately as ice grains, which flow back to the disc (right panel). In contrast, the  vapour inside the blue streamline rotates around the planet on closed streamlines in a bound atmosphere. We then define the blue streamline as the critical streamline that distinguish the recycling efficacy of H$_2$O. Most incoming pebbles already sublimate before they enter the critical streamline and do not contribute to the vapour content in the planetary atmosphere.

To further investigate the result, in \fg{coupled_contra_profile} we present azimuthally-averaged profiles of gas density, vapour fraction and temperature for runs \texttt{gasOnly} and \texttt{coupled}. The middle panel shows the radial profile of the vapour fraction. As expected, it increases sharply from the outer region and reaches the value of 0.2 for the inner atmosphere. The sharp increase denotes the sublimation front, where the ice grains sublimate and the vapour freezes out. Due to the injection of vapour the density of the \texttt{coupled} run is higher than the \texttt{gasOnly} run. There emerges a temperature plateau at the sublimation front where the temperature profile becomes flatter and lower than that in the \texttt{gasOnly} run. This plateau is due to the latent absorption associated with the phase change process, which suppress the temperature increase. Interior of the plateau where all ice is sublimated, the temperature begins to increase again at a steeper rate than the \texttt{gasOnly} run. This is because of a lower $\gamma$ value caused by vapour injection (see \eq{gamma_simplify}, \citet{BrouwersOrmel2020}).

These simulations demonstrate that the phase change module work for different hydrodynamic code and that it models thermodynamic processes like latent heat energy exchange as expected.

\subsection{Fiducial run}
\label{sec:fiducial}
For the fiducial run (\texttt{pebble-fiducial} in \tb{simu_paras}), the planet position is fixed at 4 au and the dimensionless mass $m=0.1$ (${\sim}3.3M_{\oplus}$ at 4 au). The Stokes number of the pebbles is 0.01 and the pebble-to-gas ratio is 0.01.

The flow patterns corresponding to this run at time $t=120\,\Omega^{-1}$ are shown in \fg{inject_fiducial}, when a steady state pattern has emerged. The left panel shows that pebbles enter the atmosphere where they sublimate to vapour. The streamlines clearly show the inspiralling trajectories of the accreted pebbles. To indicate how much ice is sublimated, we denote the ``ice sublimation'' band with red lines. Inside the red line, pebbles get accreted inside the sublimation front and the ice contained sublimates to vapour while outside it pebbles leave directly via the horseshoe. We note that our definition is different from the typical pebble accretion band, which denotes the accretion of refractory pebbles to the core. Because of adiabatic compression, the temperature of the ambient gas starts to increase around the planet Bondi radius. For the parameters of this simulation, the sublimation front $R_\mathrm{subl}$ -- the radius where the ice has fully sublimated from the pebbles -- happens to be around this point. Within $R_\mathrm{subl}$ all H$_2$O is in vapour form (middle panel).

Specifically, the radius of the sublimation front $R_{\mathrm{subl}}$ is defined as
\begin{equation}
    \overline{\rho_{\mathrm{ice}}}(r<R_{\mathrm{subl}}) = 0
\end{equation}
where $\rho_{\mathrm{ice}} = \rho_{\mathrm{peb,ice}}+\rho_{\mathrm{grain,ice}}$ is the total ice density and $\overline{\rho_{\mathrm{ice}}}$ is its azimuthal average. In addition, we define an atmosphere radius $R_\mathrm{atm}$ as the \textit{minimum} distance from the critical streamline to the planetary core\footnote{This definition differs from that of \citet{MoldenhauerEtal2022} where $R_{\mathrm{atm}}$ is defined based on the azimuthal velocity.}. Approximately, vapour inside $R_{\mathrm{atm}}$ is harder to be recycled back to the disc than vapour outside it. Clearly, $R_{\mathrm{subl}}>R_{\mathrm{atm}}$ in the fiducial run, which implies that the vapour can readily flow back to the disc.  We list the value of $R_{\mathrm{subl}}$ and $R_{\mathrm{atm}}$ in \tb{outputs}. 


\begin{figure}
    \includegraphics[width=\columnwidth]{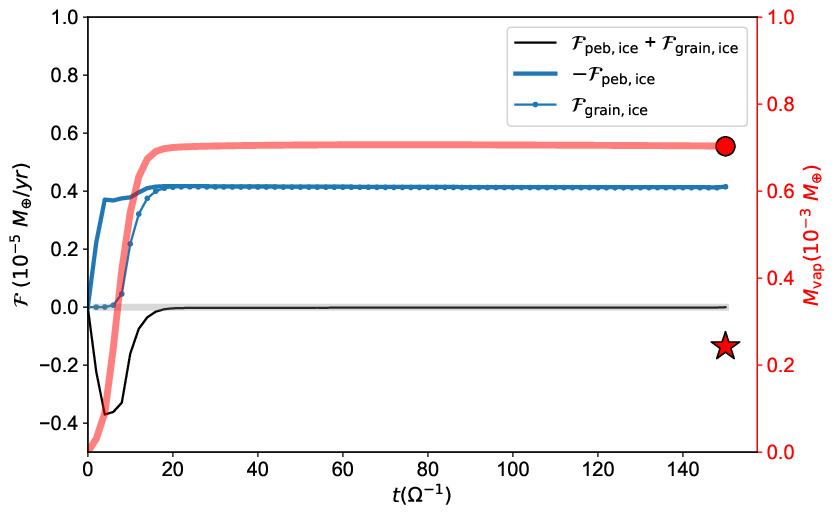}
    \caption{Mass flux and total vapour mass of run \texttt{pebble-fiducial}. The $x$-axis denotes the simulation time. The left $y$-axis represents indicates the pebble, grain, and net mass flux and the right y-axis (red) gives the total vapour mass inside the atmosphere. The final total vapour mass $M_{\rm vap}$ is marked with a red circle while the bounded vapour mass $M_{\rm vap,bound}$ (the vapour mass inside $R_\mathrm{atm}$) is marked with a star.}
    \label{fig:inject_fiducial_flux}
\end{figure}

To examine whether the atmosphere vapour content has reached a steady value, we measure the mass influx of ice contained in pebbles, $\mathcal{F}_{\mathrm{peb,ice}}$, and the outflux of recycled ice grains, $\mathcal{F}_{\mathrm{grain,ice}}$. In \fg{inject_fiducial_flux} both fluxes are measured at $3R_{\mathrm{Bondi}}$ and the sum of them gives the net ice flux. In addition, \fg{inject_fiducial_flux} presents the total vapour mass $M_{\mathrm{vap}}$ inside the whole domain (the red thick line). Not all the vapour is bounded to the planet; the vapour outside the critical streamline is on its way to freeze out as ice grains. For reference, we also give the amount of vapour inside $R_{\rm atm}$ ("star" symbol), which we define as the bound vapour mass $M_{\rm vap,bound}$  (see \fg{inject_fiducial_flux}). Initially, the net flux is negative, indicating net vapour injection. Quickly, after ${\sim}4t_{\mathrm{soft}}$, the net flux converges to zero and the total vapour mass reaches a steady value. The incoming pebbles get completely recycled back to the disc after sublimation and then freeze-out to grains. 


The steady state result is better illustrated in \fg{inject_fiducial_fraction}, where the material fractions are shown (e.g., $f_{\mathrm{peb,ice}} = \rho_{\mathrm{peb,ice}}/\rho_{\gas}$). $f_{\mathrm{satur}}$ represents the saturation profile calculated from the /local temperature and \eq{saturated}. The position of the sublimation radius as defined above is indicated by the vertical grey line. The vapour fraction $f_{\mathrm{vap}}$ reaches a steady value of ${\approx}0.013$ inside the sublimation front, which we define as $f_{\mathrm{vap,peak}}$. Here $f_{\mathrm{vap,peak}} > f_{\rm p2g,ini}/(1.0+f_{\rm p2g,ini})$, higher than the \texttt{coupled} run. This increase is a consequence of the gravitational focusing experienced by the pebbles. 


In summary, since $R_{\mathrm{subl}} > R_{\mathrm{atm}}$, incoming pebbles already sublimate before they enter the atmosphere. A steady-state emerges where vapour injection via incoming pebble sublimation is fully balanced by freeze-out of ice grains that are recycled back to the disc. The peak vapour fraction in the atmosphere $f_{\mathrm{vap,peak}}$ is higher than what is expected for well-coupled grains but remains low. 

\begin{figure}
    \includegraphics[width=\columnwidth]{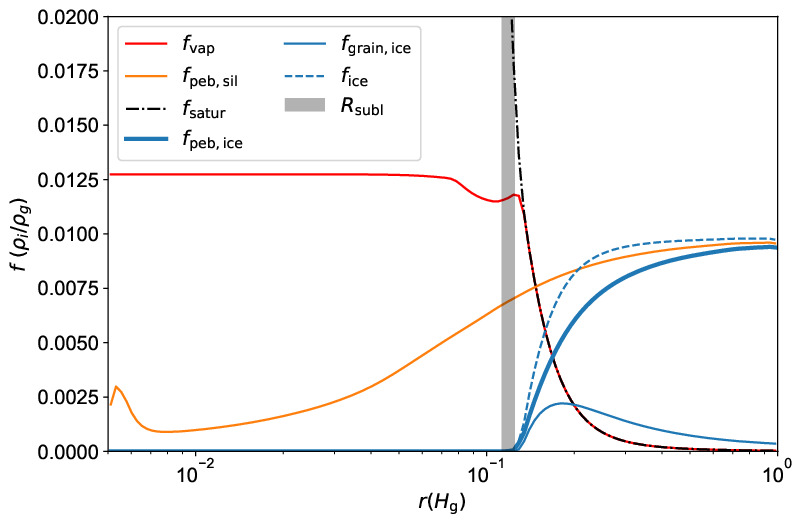}
    \caption{Azimuthally-averaged radial profiles of run \texttt{pebble-fiducial}. The $y$-axis denotes the material fraction normalized by the gas density (e.g., $f_{\mathrm{peb,ice}} = \rho_{\mathrm{peb,ice}}/\rho_{\gas}$). The black dash-dotted line represents the saturation profile and $f_{\mathrm{ice}} = f_{\mathrm{peb,ice}} +  f_{\mathrm{grain,ice}}$ denotes the total ice fraction. The silicate fraction has a small bump near the core, which is caused by the Keplerian rotation of the gas that slows down pebble infall.}
    \label{fig:inject_fiducial_fraction}
\end{figure}

\begin{figure*}
    \includegraphics[width=\textwidth]{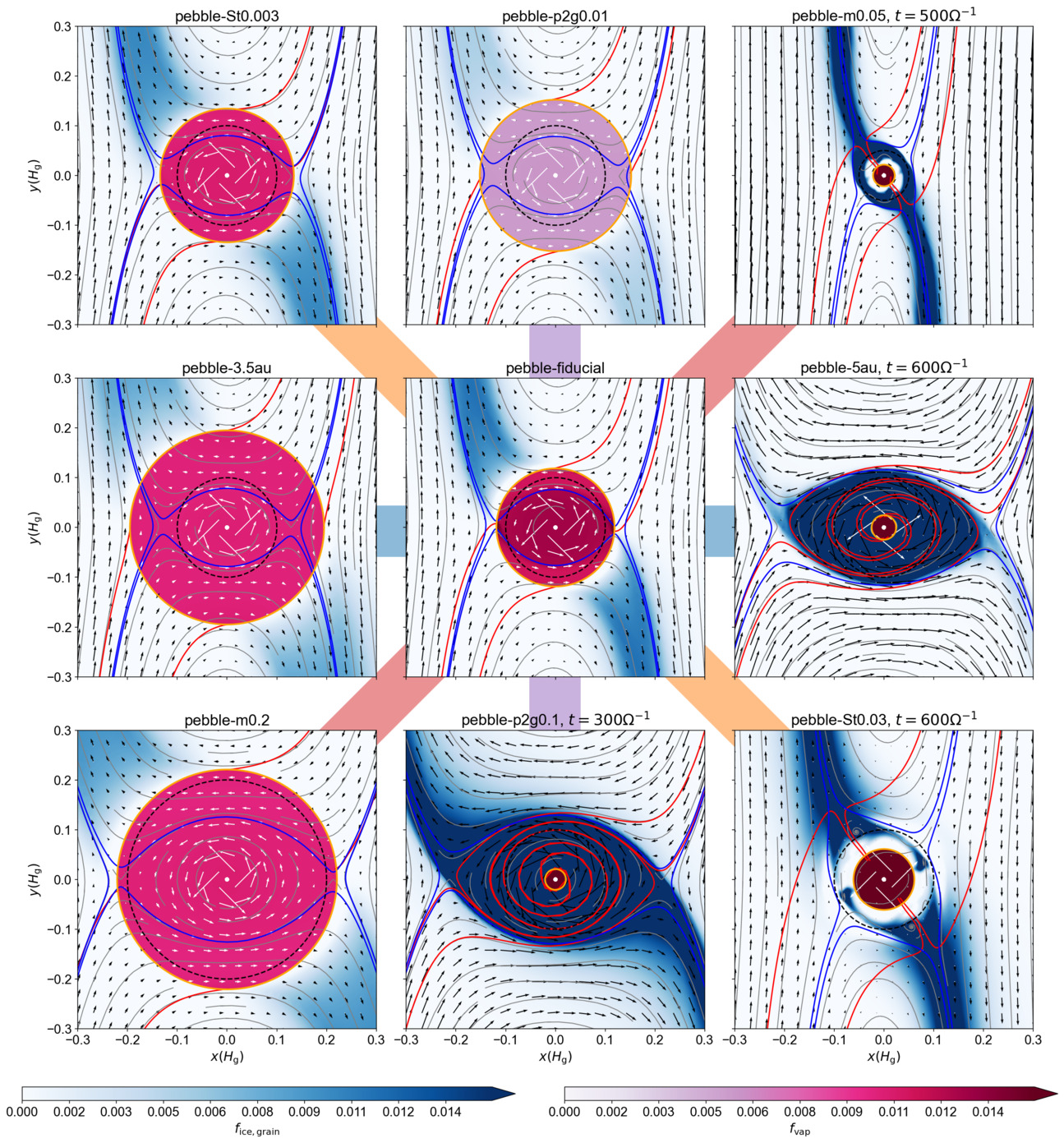}
    \caption{Flow patterns of ``pebble'' runs (listed in \tb{simu_paras}) following the simulation setup of \se{simu_set_up}. The fiducial run is shown in the centre while other panels show runs that have different parameters varied. Again the yellow circle denotes the sublimation front, the black dashed circle denotes the Bondi radius, the blue line denotes the critical streamline and the red line indicates the ice sublimation band. In each plot, the magenta-to-brown color inside the sublimation front denotes vapour fraction ($f_{\mathrm{vap}}$) while the intensity of the blue color outside it denotes ice grain fraction ($f_{\mathrm{ice,grain}}$). Movies corresponding to \texttt{pebble-3.5au} and \texttt{pebble-5au} are available as online supplementary material and on github (\href{https://github.com/rainkings/Yu_pre/blob/main/videos/supplementary/pebble-3.5au.mp4}{pebble-3.5au} and  \href{https://github.com/rainkings/Yu_pre/blob/main/videos/supplementary/pebble-5au.mp4}{pebble-5au}).}
    \label{fig:inject_patterns}
\end{figure*}

\begin{figure*}
    \includegraphics[width=0.8\textwidth]{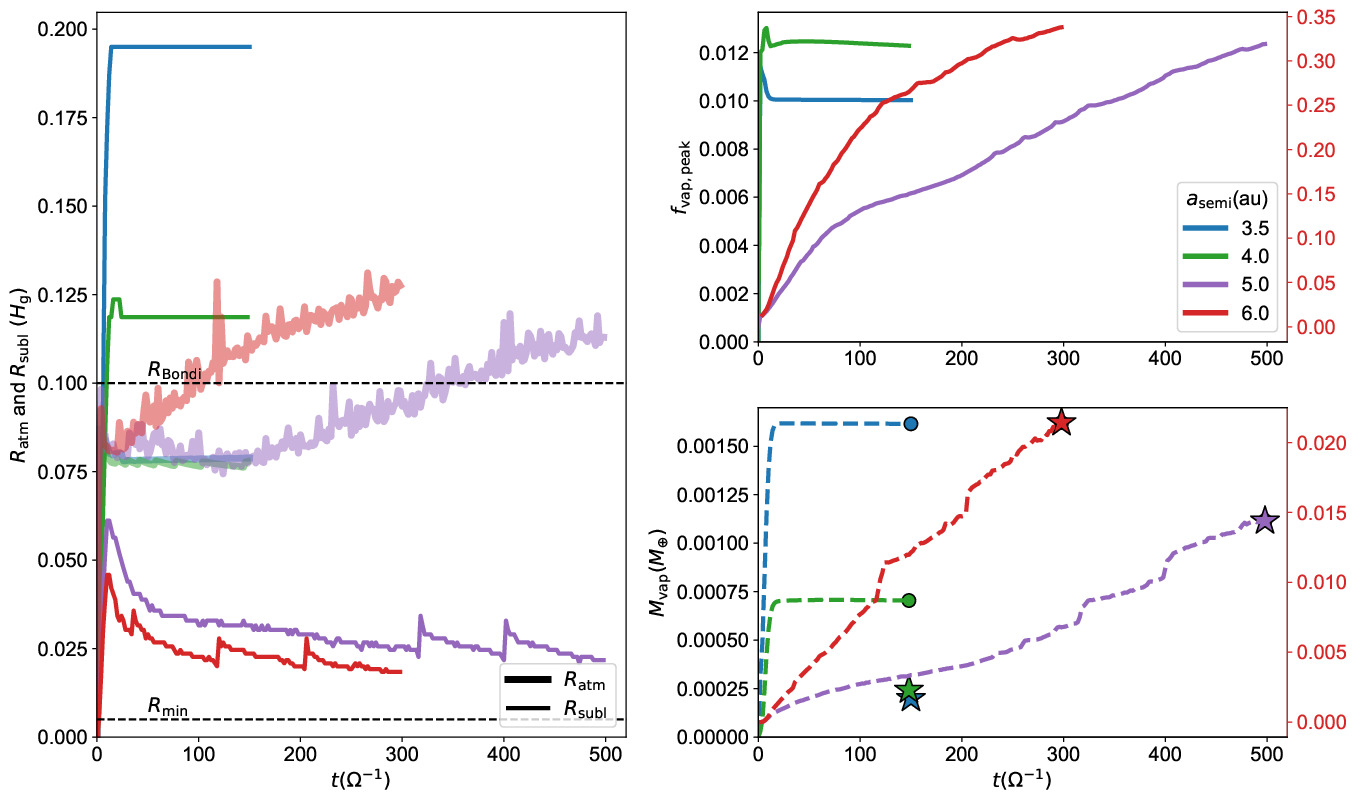}
    \caption{Characteristic outputs of runs of different location evolving with the simulation time. Left panel shows the atmosphere radius $R_{\mathrm{atm}}$ (transparent thick line) and the sublimation front $R_{\mathrm{subl}}$ (opaque thin line) together in unit of disc scale height. The Bondi radius $R_{\mathrm{Bondi}}$ and inner boundary size $R_{\mathrm{min}}$ are also indicated with black dashed lines; On the right, the upper panels shows the peak vapour fraction $f_{\mathrm{vap,peak}}$ while the lower panel vapour mass $M_{\mathrm{vap}}$. The left y-axis (black) represents the value of \texttt{pebble-3.5au} and \texttt{pebble-fiducial} while the right y-axis (red) represents \texttt{pebble-5au} and \texttt{pebble-6au} to avoid showing data that differs too much with the same axis. The dots in the last panel denote the total vapour mass contained in the domain at the final time while the stars denote the bound vapour mass.}
    \label{fig:inject_a_semi_output}
\end{figure*}

\begin{figure*}
    \includegraphics[width=\textwidth]{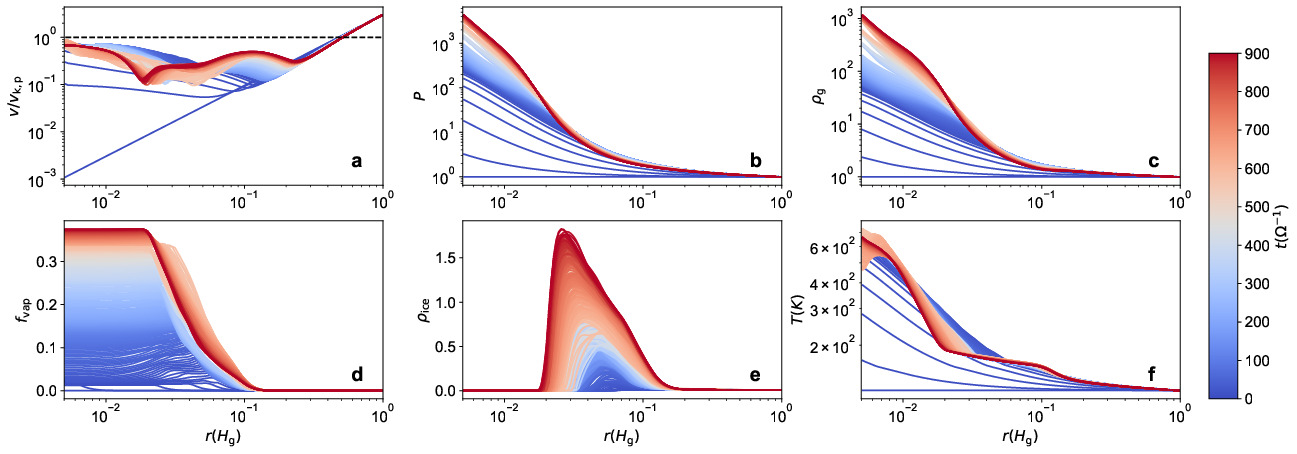}
    \caption{Time evolution of azimuthally-averaged radial profiles of run \texttt{pebble-5au}. From left to right, top to bottom, the $y$-axes show the norm of the velocity vector (normalized by the planet's Keplerian velocity), gas pressure, gas density, vapour fraction, total ice density and temperature, respectively. The $x$-axes denote distance from the planet's centre. Line color indicates simulation times, as given by the colorbar on the right. The black dashed line in the first panel represents the Keplerian limit, $v/v_{\mathrm{k}} = 1.0$.}
    \label{fig:inject_5_au_profile}
\end{figure*}

\subsection{Varying the planet position}
\label{sec:vary_a_semi}
In this section, we vary the semi-major axis of the planets while keeping other parameters the same as in the fiducial run (see \texttt{pebble-3.5au}, \texttt{pebble-5au} and \texttt{pebble-6au} in \tb{simu_paras}). \Fg{inject_patterns} is a collection of flow patterns for several parameter variations with the centre panel that of the fiducial run. In these panels we combine the vapour and ice grains fractions, which share the same velocity field. Inside the sublimation front the intensity of the color denotes the vapour fraction while outside it the color denotes the ice grain fraction. Along each of the four colored lines, one parameters is varied with respect to the fiducial run. In this section, we are concerned with the middle panels. In addition, \fg{inject_a_semi_output} presents several output quantities as function of time to quantitatively investigate the evolutionary trend.

The 3.5 au run is qualitatively similar to the fiducial run.
The time evolution of the characteristic outputs, shown in \fg{inject_a_semi_output}, reach a steady state quickly, which is similar to the default run (4 au). 
For both runs, $R_{\mathrm{subl}}>R_{\mathrm{atm}}$ and the ice contained in incoming pebbles is fully recycled by freezing-out ice grains.
At 3.5 au, however, the sublimation front is significantly larger than the Bondi radius, which is a result of the increasing ambient temperature at this planet location (\eq{disc_T}).
Also, further from the planet, pebbles experience less significant gravitational focusing by the planet and remain more tightly-coupled to the gas. This results in the drop of the peak vapour fraction for the 3.5 au run (${\approx}0.01$), which is only slightly above the value of the perfectly-coupled runs (${\approx}0.0091$). Although the 3.5 au run contains a higher total vapour mass (dashed line) than the 4 au run, less vapour is bound to the atmosphere (star, \se{fiducial}).

In contrast, the run \texttt{pebble-5au} does not reach a steady state within the simulation time (${\sim}900\,\Omega^{-1}$) and the snapshot at $t = 300~\Omega^{-1}$ is shown in \fg{inject_patterns}. Compared to the fiducial run, $R_{\rm atm}$ expands while $R_{\rm subl}$ shrinks significantly. Consequently, the sublimation front lies deep inside the atmosphere. Ice keeps being accreted (purple lines) while the vapour cannot (easily) escape the atmosphere, leading to a sustained growth of the vapour content (see \fg{inject_a_semi_output}). Although vapour does freeze-out, the ice grains similarly remain locked in the atmosphere and keep accumulating. The \texttt{pebble-6au} run (red lines) results in a picture similar to the 5 au run with an even more rapid vapour growth and higher vapour fraction.
A positive feedback loop is at work to account for these features. Initially, the sublimation fronts of the 5 and 6 au runs are already smaller than their atmosphere radii due to the lower ambient temperature (see \fg{inject_a_semi_output}). Cooling caused by ice sublimation further shrinks the sublimation front. At the same time the injection of vapour grows the atmosphere in terms of its mass, which also causes it to expands \citep{OrmelEtal2015,MoldenhauerEtal2022}. The divergence of these two critical radii suppresses the efficacy of the ice recycling. Hence, we see a continuous growth of the peak vapour fraction and vapour mass. Because of the significant expansion of the atmosphere, nearly all vapour is bound, see \fg{inject_a_semi_output}.

\begin{figure*}
    \includegraphics[width=\textwidth]{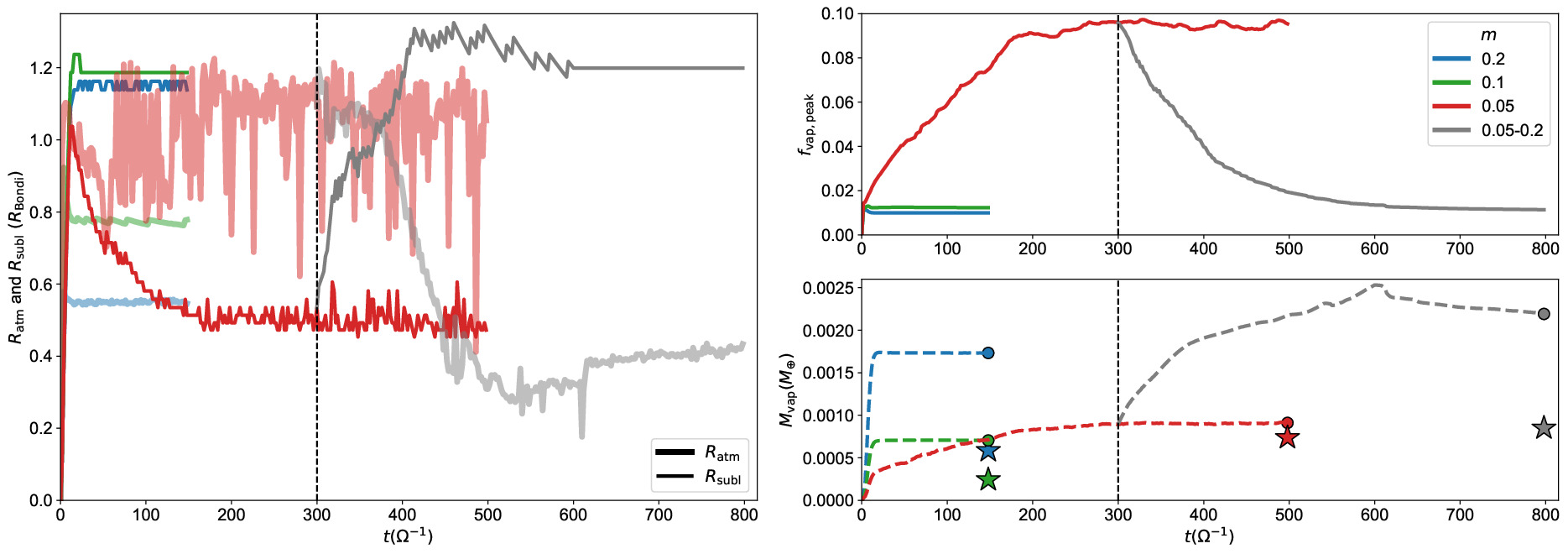}
    \caption{Characteristic outputs of runs of different planet masses evolving with the simulation time. Left and right panels are similar to \fg{inject_a_semi_output} except $R_{\mathrm{atm}}$ and $R_{\mathrm{subl}}$ are in unit of Bondi radius. The vertical dashed line in each panel denotes $t = 300 \Omega^{-1}$, which is the starting point of the run with planet mass growth from $m = 0.05$ to 0.2.}
    \label{fig:inject_m_output}
\end{figure*}
To verify this positive feedback, we plot the azimuthally-averaged radial profiles of \texttt{pebble-5au} in \fg{inject_5_au_profile}. The color of the lines indicates time in units of $\Omega^{-1}$.
First, the temperature profiles (f) keep evolving, extending the plateau (\se{coupled}) towards the core because of the continuous cooling caused by ice sublimation. This implies significant shrinking of $R_{\rm subl}$. Second, the velocity profile (a) also evolves. The radius where the azimuthally-averaged velocity displays its minimum, indicating the transition from a Keplerian shear-dominated velocity to a rotationally-dominated velocity in the planetary atmosphere, shifts outwards in time (from $r \sim 0.1$ to $0.3$ in \fg{inject_5_au_profile} a), indicating that $R_{\rm atm}$ expands. The expansion of the atmosphere is exacerbated by the decrease of the adiabatic index $\gamma$ due to the increasing vapour fraction (\eq{gamma_simplify}), which leads to more material falling in, as can also be seen in the Gas-only runs (\se{gas_only}).
Finally, the gas density (c) and pressure (b) also increase significantly. Near the core, the density increase by a factor of $\sim 15$ while the vapour fraction (d) only become $\sim 4$ times larger, meaning that the entire atmosphere becomes significantly heavier and the additional accretion of H and He contributes more than the vapour.

In summary, the respective values of $R_{\mathrm{atm}}$ and $R_{\mathrm{subl}}$ are the key parameters that determine the evolutionary trend of vapour content. When the sublimation front exceeds the (bound) atmosphere, incoming pebbles can be fully recycled and the vapour content reaches a steady value (recycling-dominated). On the other hand, when the sublimation front is inside the atmosphere, vapour tends to be locked deep in the atmosphere and keeps accumulating. The positive feedback works effectively to shrink the sublimation front and expands the atmosphere, leading to continuous growth of the vapour content (vapour-dominated) and the entire atmosphere. Consequently, the extent and the amount of vapour that a planet is able to hold on to strongly depend on its location in the disc (ambient temperature).

\subsection{Varying the planet mass}
\label{sec:vary_m}
In this section, we vary the planet mass while keeping other parameters the same as in the fiducial case (see \texttt{pebble-m0.05}, \texttt{pebble-m0.2}) in \tb{simu_paras}. The corresponding flow patterns are shown in \fg{inject_patterns} along the upper diagonal (red line). The run \texttt{pebble-m0.2} has already reached a steady state while the pattern of the \texttt{pebble-m0.05} run is still evolving and we show the snapshot at $t = 500~\Omega^{-1}$. In \fg{inject_m_output} the characteristic outputs are plotted where $R_{\rm atm}$ and $R_{\rm subl}$ are in units of their respective Bondi radius. 

The vapour fraction of the high-mass run \texttt{pebble-m0.2} quickly reaches a steady value that is slightly lower than that of the fiducial case. Initially, just after the gas infall time (at $t\sim 4 t_{\rm soft} = 20 \Omega^{-1}$), its sublimation front lies already significantly further away from the atmosphere radius $R_\mathrm{atm}$, leading to the recycling-dominated regime. We now explain the mass-dependence of $R_\mathrm{subl}/R_\mathrm{atm}$.

First, $R_{\rm subl}$ can be obtained by assuming that at the sublimation front the saturatd vapour fraction reaches $f_{\mathrm{vap,peak}}$:
\begin{equation}
    \label{eq:R_subl_quanti}
    f_{\mathrm{vap,peak}} = \frac{P_{\mathrm{eq,0}}\exp{\left(-\frac{T_{\mathrm{a},i}}{T_{\mathrm{subl}}}\right)}}{P_{\mathrm{subl}}} \frac{\mu_{\mathrm{H_{2}O}}}{\mu_{\gas}}
\end{equation}
where $P_{\rm subl}$ and $T_{\rm subl}$ are the gas pressure and temperature at the sublimation front respectively, and $\mu_{\mathrm{H_{2}O}}$ is the mean molecule weight of water. Given that the adiabatic index $\gamma$ changes little for the recycling-dominated runs, the temperature and pressure profile follows the 1D adiabatic solution:
\begin{equation}
    \label{eq:adiabatic_solution}
    \begin{aligned}
        T_{\rm subl} &= \left[1+\frac{\gamma-1}{\gamma} R_{\rm Bondi}\left(\frac{1}{R_{\rm subl}}-\frac{1}{r_{\rm max}}\right) \right]T_{\rm disc} \\
        P_{\rm subl} &= \left[1+\frac{\gamma-1}{\gamma} R_{\rm Bondi}\left(\frac{1}{R_{\rm subl}}-\frac{1}{r_{\rm max}}\right) \right]^{\frac{\gamma}{\gamma-1}}P_{\rm disc}    
    \end{aligned}
\end{equation}
where $T_{\rm disc}$ and $P_{\rm disc}$ are the disc background temperature and pressure and $r_{\mathrm{max}}$ denotes outer boundary of the simulation domain.
From \eqs{R_subl_quanti}{adiabatic_solution}, $R_{\rm subl}/R_{\rm Bondi}$ is solely determined by $f_{\mathrm{vap,peak}}$ given disc parameters. Initially, within the gas infall time (${\sim}4 t_{\rm soft}$), all runs shown in \fg{inject_m_output} have a vapour fraction similar to their initial pebble-to-gas ratio ($\sim 0.01$) since H$_2$O vapour has not yet accumulated significantly. Hence, they share a similar $R_{\rm subl}/R_{\rm Bondi} \sim 1.2$.

\begin{figure*}
    \includegraphics[width=0.8\textwidth]{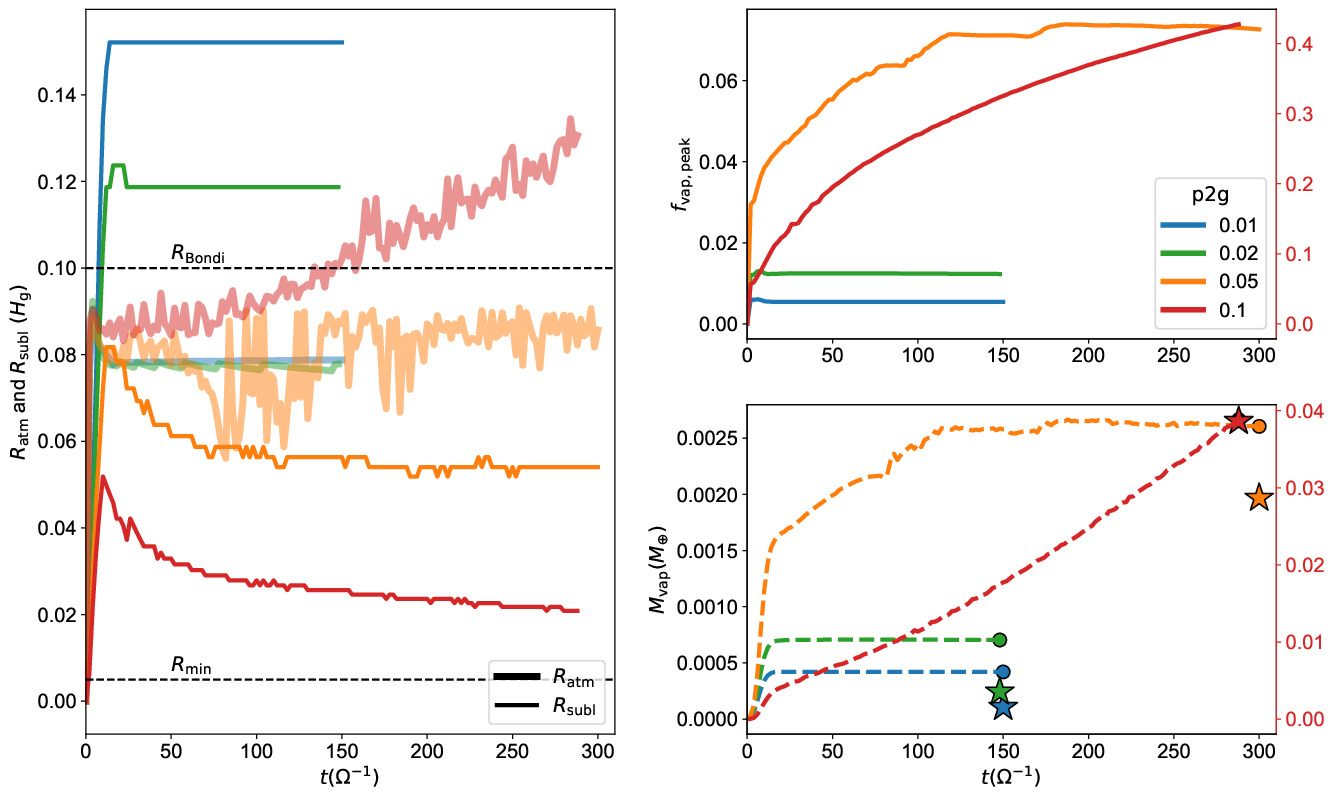}
    \caption{Characteristic outputs of runs of different $f_{\mathrm{p2g,ini}}$ evolving with the simulation time. Left panel is like in \fg{inject_a_semi_output}; Right panel is similar to that in \fg{inject_a_semi_output} while the right y-axis (red) represents \texttt{pebble-p2g0.1}. }
    \label{fig:inject_p2g_output}
\end{figure*}

On the other hand, $R_{\rm atm}/R_{\rm Bondi}$ shows a clear trend with mass. Initially (at $t \sim 4t_{\rm soft}$), $R_{\rm atm}/R_{\rm Bondi}$ is smaller with increasing planet mass (\fg{inject_m_output}). We can understand this behavior in terms of the mass-dependence of the horseshoe width. It has been shown that, when ignoring atmospheric feedback,\footnote{If the atmosphere becomes very massive (by vapour injection e.g.), it tends to growth in size as well (at least in 2D) \citep{OrmelEtal2015,MoldenhauerEtal2022}, as is also seen in this work.} the half-width of the widest horseshoe orbit scales as $x_{\rm hs} \propto \sqrt{m}$ \citep{PaardekooperPapaloizou2009,Ormel2013,MassetBenitez-Llambay2016}. The corresponding streamline -- the critical streamline -- passes through the stagnation point at $(\pm x_\mathrm{sep},0)$. For barotropic flows, the Bernoulli constant $B$
\begin{equation}
    \label{eq:Bernuolli}
    B = \frac{1}{2}v_{\gas}^{2} + W + \Phi_{\rm p} +\omega_\infty\Omega x^{2}
\end{equation}
is conserved along streamlines \citep{Ormel2013}, where $W$ is the enthalpy, $\Phi_{\rm p}$ is the planet potential and $\omega_{\infty} = -\frac{3}{2}\Omega$ is the vorticity in the far field (far from the planet). Under the adiabatic EOS where $P = \rho_{\gas}^{\gamma}$ (in dimensionless units, see \se{disc_model}), the enthalpy can be calculated as,
\begin{equation}
    \label{eq:entralpy}
    W = \int \frac{\mathrm{d} P}{\rho_{\gas}} 
    = \int \frac{P}{\rho_{\gas}} \frac{\mathrm{d} \log P}{\mathrm{d} r} \mathrm{d}r
    = \frac{m}{r}
\end{equation}
after substituting the adiabatic solution for a hydrostatically-supported atmosphere, \eq{entralpy}.
Therefore, $\Phi_\mathrm{P} + W \approx 0$ at the separatix location as well as in the far field. Equating the Bernoulli expression in the far field and at the separatix, respectively, gives:
\begin{equation}
    \frac{1}{2} (\frac{3}{2} x_\mathrm{hs})^2 -\frac{3}{2} x_\mathrm{hs}^2 = -\frac{3}{2} x_\mathrm{sep}^2.
\end{equation}
Hence, $x_{\rm sep} \propto x_\mathrm{hs} \propto \sqrt{m}$. And $x_{\rm sep}$ is half of the distance of two stagnation points, which denotes the long axis of the atmosphere ($R_{\rm atm}$ denotes the short axis of the atmosphere). If we further assume that $x_\mathrm{sep}$ scales with $R_\mathrm{atm}$, then $R_{\rm atm}/R_{\rm Bondi} \propto m^{-1/2}$ accordingly.
This means that with increasing planet mass $m$, $R_{\rm atm}/R_{\rm Bondi}$ will decrease, which explains the trend of initial values of $R_{\rm subl}/R_{\rm atm}$ as shown in \fg{inject_m_output}. Furthermore, the gravity at the sublimation front is $-m/R_{\rm subl} \propto m^{-1}$. With a larger planet mass, pebbles go through a smaller gravitational focusing near the sublimation front, leading to the behavior closer to the perfectly-coupled runs, which also implies a vapour fraction similar to the perfectly-coupled runs. Therefore, the run \texttt{pebble-m0.2} has a lower peak vapour fraction compared with the fiducial run (\fg{inject_m_output}). 

For the run \texttt{pebble-m0.05}, its initial $R_{\rm atm}/R_{\rm Bondi}$ becomes higher, which reaches the sublimation radius and, $R_{\mathrm{subl}} \approx R_{\mathrm{atm}}$. Instead of quickly reaching a steady state (recycling-dominated) or entering rapid vapour growth by the positive feedback, it instead reaches a quasi-steady state where all the output quantities evolve slowly due to the comparable rate of sublimation and freeze-out. This ``intermediate'' regime has its $R_{\rm subl}$ gradually shrink due to the cooling from ice sublimation while $R_{\rm atm}$ keeps fluctuating around a value of 1.1$R_\mathrm{Bondi}$. Compared to higher-mass runs, it has significantly higher vapour amounts in the atmosphere, but it is still not high enough to expand the atmosphere (the vapour fraction reaches ${\sim}0.2$ before $R_\mathrm{atm}$ begins to expand in \texttt{pebble-5au}). Consequently, the atmosphere does not become significantly heavier, in contrast to the vapour-dominated runs (\se{vary_a_semi}). For its flow pattern, near the Bondi radius (\fg{inject_patterns}), there is a region where the velocity becomes very low (the arrows almost vanish).  Due to this low velocity, the critical streamline is not well defined, causing $R_\mathrm{atm}$ to fluctuate strongly and the horseshoe orbit to become narrower. As a result, the gas near the core seems to be isolated from the horseshoe orbit.

\begin{figure*}
    \includegraphics[width=0.8\textwidth]{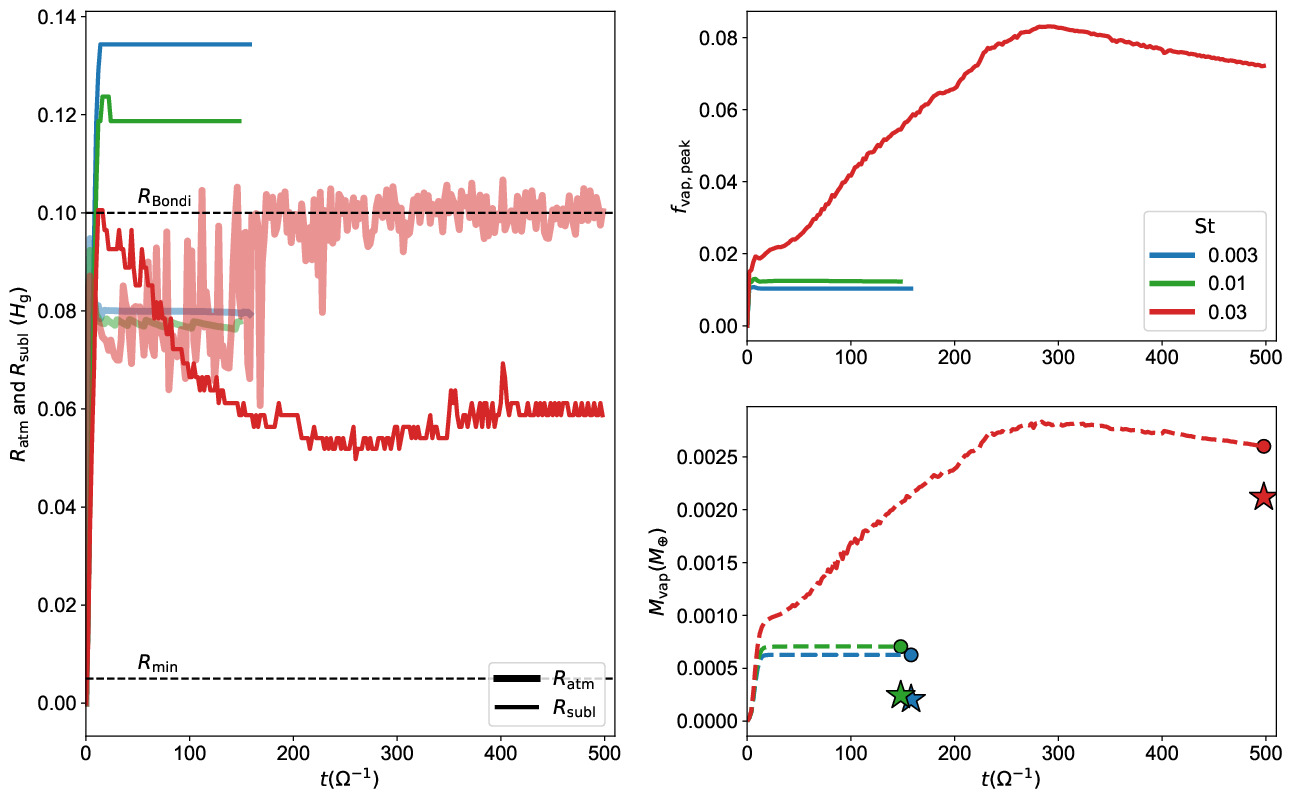}
    \caption{Characteristic outputs of runs of different $\mathrm{St}$ evolving with the simulation time. Similar to \fg{inject_a_semi_output} with only one y-axis for the right panels.}
    \label{fig:inject_St_output}
\end{figure*}
In order to find the reason for the low amount of circulation, we inspect the time evolution of this run. Initially, the flow pattern is similar to the recycling-dominated cases. However, a positive radial gradient of the vapour fraction emerges due to vapour piling up near the sublimation front. In addition, an azimuthal imbalance (both in temperature and pressure) also emerges caused due to ice preferentially sublimating near the separatix points. This triggers a strong compositional instability (see \se{caveats} for more discussions), which transports the piling-up vapour towards the core and excites pressure waves that reshape the flow pattern. In contrast, this instability does not happen in vapour-dominated runs because the atmosphere already expand significantly and the vapour pile-up happens deep inside the atmosphere. It also does not happen in the recycling-dominated runs as, there is no significant vapour pile-up. After this instability, the flows at the horseshoe orbit seem to experience additional pressure support contributed by the vapour, which weakens the horseshoe orbit.
The high vapour fraction does not weaken the horseshoe orbit in the vapour-dominated runs, because the sublimation front resides deep inside the atmosphere, far from the horseshoe orbit (\fg{inject_patterns}, \texttt{pebble-5au} e.g.). The additional pressure support contributed by vapour is tentatively seen in \fg{inject_5_au_profile} a, where there is a dip of velocity profile near the sublimation front.

The fact that the vapour fraction drops with increasing planet mass may seem counterintuitive. We also observe this phenomenon for planets at 4.5au (see \tb{outputs}) where the \texttt{pebble-4.5au} is classified as intermediate and the higher mass \texttt{pebble-m0.2-4.5au} as recycling-dominated.
To further demonstrate this point, we run a simulation starting from $t = 300 \Omega^{-1}$ of the vapour-dominated \texttt{pebble-m0.05} run and let the planetary core grow linearly within $300\,\Omega^{-1}$ to $m =0.2$. In \fg{inject_m_output} (grey line), the peak vapour fraction drops significantly and finally relaxes to the same value of \texttt{pebble-m0.2}. Similarly, the total vapour mass first increases with $R_{\mathrm{subl}}$ expanding sharply due to gravitational energy release. Then, after the core mass reaches $m = 0.2$ ($t = 600 \Omega^{-1}$), $M_{\mathrm{vap}}$ begins to decrease, with recycling flows gradually removing vapour from the atmosphere.

In summary, 
we found that with increasing mass, planet's atmospheres, while becoming bigger, may hold less (bound) vapour and that (sometimes) our qualitative assessment -- recycling, intermediate, or vapour-dominated -- may shift towards the former. The reason is that, although $R_\mathrm{atm}$ increases with $m$, the sublimation radius increases faster.

\subsection{Varying the pebble-to-gas ratio}
\label{sec:vary_p2g}
Next, we vary the initial pebble-to-gas ratio while other parameters are kept the same as in the fiducial run (see \texttt{pebble-p2gxxx} in \tb{simu_paras}). The flow patterns of \texttt{pebble-p2g0.01} and \texttt{pebble-p2g0.1} are shown in \fg{inject_patterns} along the purple line. The run \texttt{pebble-p2g0.1} is still evolving by the time the snapshots is taken ($t = 300~\Omega^{-1}$). \Fg{inject_p2g_output} presents the time evolution of the characteristic output quantities.

Varying the pebble-to-gas ratio alters the pebble flux. With a higher pebble flux, more ice sublimates per unit time and more energy (latent heat) is absorbed from the environment. This mitigates the increase of the temperature caused by the gravitational energy release. Therefore, higher $f_{\mathrm{p2g,ini}}$ results in smaller sublimation radii initially (\fg{inject_p2g_output}, $t \sim 4 t_{\rm soft}$). For the run with $10\%$ pebble-to-gas (\texttt{pebble-p2g0.1}), the positive feedback effect (\se{vary_a_semi}) results in continual atmosphere expansion and simultaneous shrinkage of the sublimation front. On the other hand, for the low pebble-to-gas ratio runs (\texttt{pebble-p2g0.01} and \texttt{pebble-p2g0.02}), $R_{\mathrm{subl}} > R_{\mathrm{atm}}$ initially. Ice contained in pebbles is fully recycled and the vapour content reaches a steady value -- the recycling-dominated limit (see \se{vary_a_semi}).

For intermediate values of the pebble-to-gas ratio (0.05), $R_{\mathrm{subl}} \approx R_{\mathrm{atm}}$ initially ($t \sim 4 t_{\rm soft}$) and the intermediate regimes emerges, which has its vapour content slowly grow and $R_\mathrm{atm}$ fluctuate more widely. Its flow pattern (not shown in \fg{inject_patterns}) is also characterized by weak amount of circulation near the horseshoe and isolated inner atmosphere, similar to the $m=0.05$ run described above.

In summary, the evolutionary trend of the vapour content is again determined by the respective values of $R_{\mathrm{atm}}$ and $R_{\mathrm{subl}}$. However, instead of shrinking the sublimation front directly by moving the planet to colder regions (\se{vary_a_semi}), increasing the pebble-to-gas ratio shrinks the sublimation front as a higher rate of sublimation cools the atmosphere. By varying the pebble-to-gas ratio (from 0.01 to 0.1), we observe a complete transition from recycling-dominated flows to the intermediate and finally the vapour-dominated regimes.

\subsection{Varying the Stokes number}
\label{sec:vary_St}
The default Stokes number $\mathrm{St} = 0.01$ represents $\mathrm{cm}$-sized ($\approx$ 0.9 cm) pebbles at 4 au. We vary them by one magnitude ($\mathrm{St}=$0.003 and 0.03) (see \fgs{inject_patterns}{inject_St_output}). The effect of varying the Stokes number is similar to varying the pebble-to-gas ratio, both of which will alter the pebble flux. In figure \fg{inject_St_output}, the initial $R_{\rm subl}$ is smaller with higher Stokes number, which is caused by the higher pebble flux. Pebbles with smaller Stokes number are more easily recycled and the recycling efficacy will be the maximum for the perfectly-coupled pebbles ($\mathrm{St}=0$ particles), which sets the lower limit of the peak vapour fraction. 

For higher Stokes number runs (\texttt{pebble-St0.03}), the intermediate regime emerges with its characteristic weak horseshoe orbits (\fg{inject_patterns}). Although $R_{\mathrm{atm}}$ shows significant expansion, the ratio of the $R_{\rm subl}$ and $R_{\rm atm}$ is not that extreme compared to vapour-dominated runs (\fgs{inject_a_semi_output}{inject_p2g_output}), thus the positive feedback does not emerge to promote further growth of vapour and expand atmosphere. Its sublimation front expands gradually after $t \approx 300~\Omega^{-1}$, showing a converging trend of $R_{\rm atm}$ and $R_{\rm subl}$ at late times, which causes the vapour content to decrease. This process is similar to the positive feedback discussed in \se{vary_a_semi} but in the opposite direction: net ice freezes-out heats the sublimation front region, causing it to expands. However, the vapour fraction is still too small to alter $R_{\rm atm}$ significantly, rendering this process much weaker.

\section{Discussion}
\label{sec:discussion}

\subsection{The composition of planets and atmospheres}
\label{sec:composition}
\begin{figure}
    \centering
    \includegraphics[width=0.9\columnwidth]{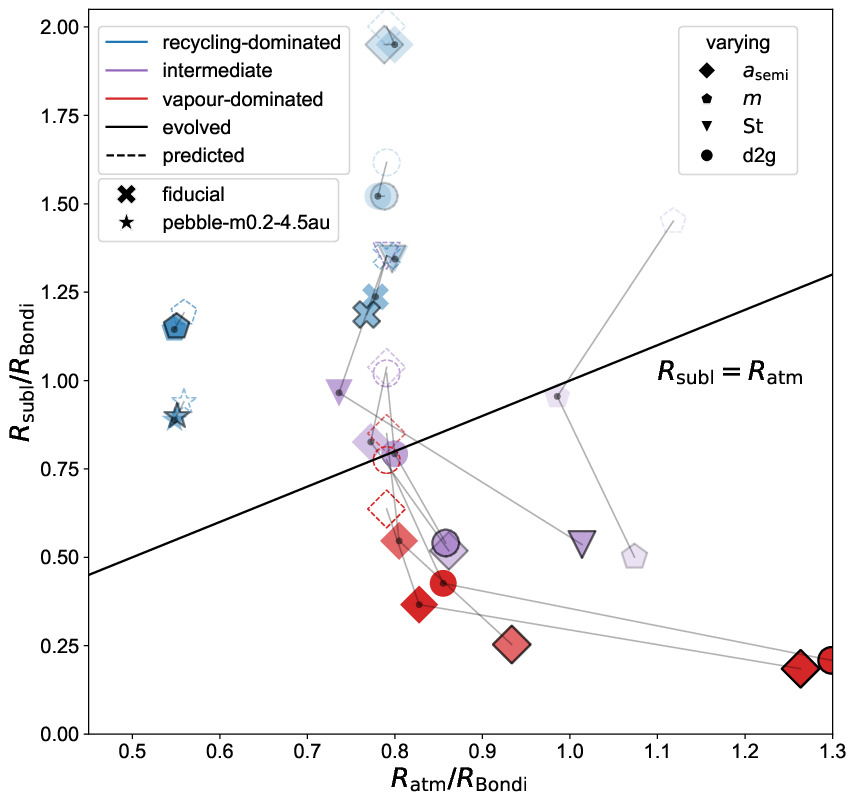}
    \caption{Summary of the characteristic outpus of all runs on the ($R_{\mathrm{atm}}/R_{\mathrm{Bondi}}$, $R_{\mathrm{subl}}/R_{\mathrm{Bondi}}$) plane. The grey lines connect different outputs of the same run with the filled, edge-free symbols denoting the outputs at $t = 4t_{\rm soft}$ (initial), the filled, solid-edged symbols denoting the outputs after evolving time ($t_{\mathrm{evolv}}$) and the open, dashed-edged symbols denoting the predicted value for perfectly-coupled grains (\se{composition}). Black nodes are also put on the initial outputs for highlighting. Different symbols denote which parameters are varied with respect to the fiducial run (see \tb{simu_paras}) with the opacity indicating different values in the sequence. For example, increasing the opacity of the circles indicates increasing dust-to-gas ratios (0.05, 0.1, and 0.2). The fiducial run is indicated by the cross and run \texttt{pebble-m0.2-4.5au} by a star.}
    \label{fig:scaling}
\end{figure}

In \fg{scaling} we summarize all runs listed in \tb{simu_paras} on the ($R_{\mathrm{atm}}/R_{\mathrm{Bondi}}$, $R_{\mathrm{subl}}/R_{\mathrm{Bondi}}$) plane. Symbols represent a certain parameter while the opacity of the color indicates the values that are varied for this parameter. The run \texttt{pebble-m0.2-4.5au}, where two parameters are varied with respect to the default, is indicated by the star. For each run three kind of symbols are plotted: open, dashed edge; filled, no edge; and filled, solid edge. The filled, edge-free symbols (also with black nodes inside) represent the initial state of the runs ($t = 4t_{\rm soft}$), which are clearly separated by the line $R_{\rm subl} = R_{\rm atm}$: the recycling-dominated runs (blue; above the line), the intermediate runs (purple; around the line) and the vapour-dominated runs (red; below the line).
The evolved state (filled, solid edge) plots the outputs after time $t_{\mathrm{evolv}}$ (\tb{outputs}, $150\ \Omega^{-1}$ for the recycling-dominated runs and $300\ \Omega^{-1}$ for the intermediate and vapour-dominated runs). For the recycling-dominated runs, both $R_{\rm subl}$ and $R_{\rm atm}$ hardly change and the symbols tend to overlap. On the other hand, for the vapour-dominated runs, the symbols have $R_{\mathrm{atm}}$ and $R_{\mathrm{subl}}$ separate divergently ($R_\mathrm{atm}$ increases, $R_\mathrm{subl}$ decreases) through the positive feedback mechanism explained in \se{vary_a_semi}. The intermediate runs move away from the $R_{\rm subl} = R_{\rm atm}$ line mostly by the decrease of $R_{\mathrm{subl}}$ while $R_{\mathrm{atm}}$ expands less significantly due to the limited accumulation of vapour (see \se{vary_St} for details).

Finally, the open, dashed-edged symbols in \fg{scaling} indicate our ``predicted'' values for $R_\mathrm{subl}$ and $R_\mathrm{atm}$.  In \se{vary_m}, we already derived $R_{\rm atm}/R_{\rm Bondi} \propto m^{-1/2}$ and we fit the prefactor as,
\begin{equation}
    \frac{R_{\rm atm}}{R_{\rm Bondi}} = 0.25 m^{-1/2}.
\end{equation}
Assuming that sublimation happens in a very narrow radial range and the adiabatic solution still fits well (\eqs{R_subl_quanti}{adiabatic_solution}), $R_\mathrm{subl}$ can be numerically solved given $f_{\mathrm{vap,peak}}$, disc parameters ($T_{\mathrm{disc}}$,$P_{\mathrm{disc}}$) and material constants (see \tb{sublimation_const} and \app{estimate}).
Here we take $f_{\rm vap,peak} \approx f_{\rm ice,ini}$, where $f_{\rm ice,ini}$ is the initial ice fraction ($f_{\rm ice,ini} = f_{\rm p2g,ini} f_{z}$, see \se{coupled} and \tb{simu_paras}). This approximation is of course only appropriate for the tightly coupled runs. Also, we assume $\gamma$ can be taken a constant.
The prediction for recycling-dominated runs fits the simulation results well, because the vapour fraction stays low. For intermediate and vapour-dominated runs, the prediction tends to overestimate $R_{\rm subl}$ because the vapour fraction at $t = 4t_{\rm soft}$ is already much higher than that in the perfectly-coupled runs, which also causes the temperature profile to deviate from the adiabatic solution of pure H-He gas (\eq{adiabatic_solution}). 

\begin{figure}
    \centering
    \includegraphics[width=\columnwidth]{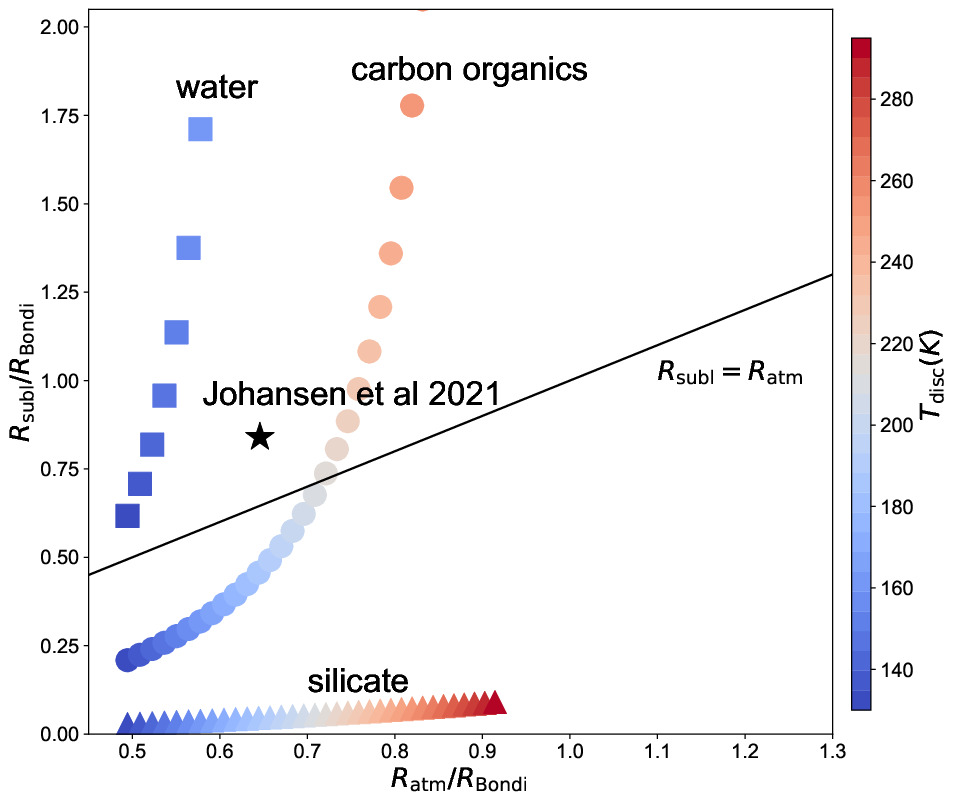}
    \caption{Predictions of $R_{\mathrm{atm}}$ and $R_{\mathrm{subl}}$ for different disc temperatures and materials of interest. Colored symbols denote different materials labelled in the figure while the black star represents water ice under the disc conditions of \citet{JohansenEtal2021} at $t = 5\,\mathrm{Myrs}$. The black solid line denotes $R_{\mathrm{subl}} = R_{\mathrm{atm}}$.}
    \label{fig:prediction}
\end{figure}
While we could improve the prediction by correcting it for, e.g., the Stokes number, the trend is never the less clear. Conditions where the predicted value yields $R_\mathrm{subl}/R_\mathrm{atm} \gg 1$ will turn out to become recycling-dominated, while for $R_\mathrm{subl}/R_\mathrm{atm} \ll 1$ the planet becomes vapour-dominated. We expect this result to apply for different disc conditions and for different materials.

A key application is that of water delivery to Earth. With the waning of the viscous accretion rate as well as the decline of the Sun's luminosity on its descent to the pre-main sequence, it is believed that the disc H$_2$O iceline would have crossed Earth's orbit \citep{GaraudLin2007,OkaEtal2011,BaraffeEtal2015,IdaEtal2016}. This would have allowed proto-Earth to accrete H$_2$O ice-rich pebbles and turn wet, inconsistent with its low water content ($2000-3000 \mathrm{ppm}$ \citealt{Marty2012}).
Recently, \citet{JohansenEtal2021} argued that recycling flows prevented accumulation of too much water. In \fg{prediction} the point corresponding to the conditions in their simulations \footnote{We adopt similar disc parameters ($T_{\mathrm{disc}} = 130 \mathrm{K}$, $\rho_{\mathrm{disc}} \approx 10^{-11} \si{g~cm^{-3}}$, $f_{\mathrm{ice,ini}} \approx 0.0013$) at $5 \mathrm{Myrs}$ for a proto-earth ($\sim 0.6 M_{\oplus}$, before the giant impact to produce moon) at 1 au, which is just before the disc dissipation time assumed in \citet{JohansenEtal2021}.}  is shown by the black star. 
As this point lies above the $R_{\mathrm{subl}} = R_{\mathrm{atm}}$ line, a recycling-dominated atmosphere is a possibility. 
If recycling-dominated or intermediate, we expect the vapour contents in the atmosphere to stabilize and even at a level of ${\lesssim}10 \%$ (characteristic values for the intermediate regime, see \tb{outputs}), the contribution to the bulk composition would still be tiny (${\sim}10 \mathrm{ppm}$), given that the characteristic envelope mass of terrestrial planets is ${\sim}10^{-4} M_{\oplus}$ \citep{JohansenEtal2021}.  
However, a transition to an H$_2$O vapour-dominated envelope (and planet) -- inconsistent with Earth-- can also not be ruled based on our results. 

The estimator can also be applied towards materials other than water. Based on the disc model adopted in this work (\se{disc_model}), we investigate retention of carbon and silicates of an accreting Earth-mass planet at 1 au. In our solar system, the main carbon carrier that contributes ${\sim}80\%$ of the total amount in solids are complex organics which sublimate at ${\sim}325-425\,\mathrm{K}$ \citep{GailTrieloff2017}.
As a proxy, we take the properties of pyrene, one of the common Polycyclic Aromatic Hydrocarbons (PAHs), (see \tb{sublimation_const}). 
We further assume that pebbles roughly contain $5 \%$ of their mass in carbon, similar to the CI chondrites \citep{JohansenEtal2021}. 
For the silicate, we take olivine as the proxy \citep{KingEtal2015} and assume that it accounts for $50 \%$ of the pebble mass. The pebble-to-gas ratio is set as in the fiducial run. 
To account for the changing luminosity of the Sun, we vary the disc temperature from 130 to 300\,K (see \fg{prediction}) and the disc pressure accordingly. 
With $T_{\mathrm{disc}}$ increasing, $R_{\mathrm{atm}}/R_{\mathrm{Bondi}}$ increases due to the increase of local thermal mass, which lowers $m$ with the same planet mass. Except for the lowest disk temperatures, water will be recycled, resulting in a dry planet. To retain the carbon, $T_{\mathrm{disc}}$ needs to significantly decrease (below ${\sim}200\,\mathrm{K}$). Hence, Earth's low-carbon fraction \citep{Marty2012} may indicate the disc stayed above this temperature. In contrast, silicates always fall in the vapour-dominated regime, implying injection of silicate vapour deep into the atmosphere. This ``vapour-planet'' scenario has already been investigated by 1D analytical \citep{BrouwersOrmel2020} and numerical \citep{OrmelEtal2021} modelling highlighting the formation of a ${\sim}1\,M_\oplus$, vapour-rich atmosphere for sub-Neptune planets. Our 2D work suggest that silicate recycling flows are ineffective to alter these findings.

\subsection{Caveats}
\label{sec:caveats}
Our pioneering work has adopted several simplifications which may affect our results: simulations were conducted in 2D without accounting for a disc headwind, did neither include self gravity nor radiation transport (RT), and did not consider frictional heating of the gas by the accreted pebbles (dust feedback). The omission of the latter two compensate each other to some extent, such that a (quasi-)steady state is often still reached. In our simulation ommiting the frictional heating, higher pebble flux leads to the vapor-dominated regime because the cooling by ice sublimation pushes the $R_{\mathrm{subl}}$ inwards (as illustrated in \se{vary_p2g} \& \ref{sec:vary_St}). However, in reality, higher pebble flux will increase the temperature of the atmosphere by depositing accretion energy of refractories throughout the atmosphere \citep{PollackEtal1996,Rafikov2006,ZhuEtal2021}. This would probably render the trend in the opposite direction: with higher pebble flux, $R_{\mathrm{subl}}$ moves outwards and the atmosphere ends up in the recycling-dominated regime. With RT included, the outer regions of pre-planetary atmospheres are likely to be (nearly-)isothermal on top of an optically thick interior -- either radiative or convective \citep{PollackEtal1996,InabaIkoma2003,Rafikov2006,OrmelKobayashi2012,MordasiniEtal2012}. The presence of an isothermal outer layer, therefore, would shift the $R_{\mathrm{subl}}$ inwards, by an amount that depends on the opacity of the gas or that of the pebbles and recondensed ices \citep{OrmelEtal2021,BrouwersEtal2021}. In contrast, the headwind induced by the disc pressure gradient could possibly shift $R_{\mathrm{atm}}$ inwards, depending on the disc structure and planet mass \citep{Ormel2013,OrmelEtal2015i,KurokawaTanigawa2018,MoldenhauerEtal2022}.

Including RT may affect the flow instability seen in the ``intermediate'' runs. In these simulations ice preferentially sublimated near the separatix points, which resulted in an azimuthal imbalance (in temperature and pressure). It needs to be verified whether the ensuing instability will also appear in simulation that include RT or whether the simulations presently classified as ``intermediate'' will evolve into either the recycling- or vapour-dominated regimes. Interestingly, the interior region in these simulations reveals a weak horseshoe motion similar to that of the $\beta$-cooling approach adopted by \citet{KurokawaTanigawa2018}, suggesting that this low-entropy material is hydrodynamically shielded from the recycling-dominated flows -- the buoyancy barrier. In our case, the entropy gradient caused by ice sublimation possibly accounts for the buoyancy barrier. But here, again, RT may act to at least partially erase these features \citep{MoldenhauerEtal2022}. In addition, including RT is likely to ameliorate the entropy violation issue that we occasionally witnessed in our simulations (\se{gas_only} and \ref{sec:fiducial}).

Three-dimensional (3D) flows are known to be qualitatively different from their 2D counterparts. In 3D, material tends to flow in through the polar region and flow out in the midplane (equatorial flows; \citealt{OrmelEtal2015i,FungEtal2015,SzulagyiEtal2016,CimermanEtal2017,MoldenhauerEtal2022}). In 2D, simulations tend to be characterized by large rotational flows around the core  (\citealt{OrmelEtal2015,BethuneRafikov2019} or see e.g.\ \fg{inject_5_au_profile}), whereas the amount of rotation in 3D simulations is much weaker due to the polar inflow \citep{OrmelEtal2015i, CimermanEtal2017} at least for low-mass planet and non-isothermal EOS \citep{FungEtal2019}. This may lead to more efficient mixing, smoothing out the radial gradient of vapour fraction, which again questions the emerge of instabilities seen in the intermediate runs. In 3D, \citet{MoldenhauerEtal2021} showed a fully-recycled atmosphere on a timescale of ${\sim}10^{3}$ orbits. But whether the same picture emerges in 3D for sublimated pebbles is still to be investigated. We envision a similar dichotomy as seen in this work with planets accreting modestly volatile pebbles at low rates attaining a steady state and operating in the recycling limit, whereas more refractory materials will be trapped in the atmosphere ($R_\mathrm{subl}/R_\mathrm{atm}\ll1$).


Another key assumption we adopted is that recondensed ice grains flow back to the disc together with the gas. However, solid recycling flows may be arrested when the condensed particles become too large to settle back to the planet. Whether or not grains will leave the atmosphere depends on how fast ice grains grow into larger particles. The coagulation timescale of  grains is \citep{OrmelEtal2007,KrijtEtal2016,JohansenEtal2021}
\begin{equation}
    \tau_{\rm coag} \sim \frac{s_{\rm ice} \rho_{\bullet,\mathrm{ice}}}{\Delta v \rho_{\rm ice}}
\end{equation}
where $s_{\rm ice}$ is the radius of ice grains, $\rho_{\bullet,\mathrm{ice}}$ is its internal density, $\Delta v$ the particle collision speed and $\rho_{\rm ice}$ the mass density of ice grains. Assuming that the relative velocity is dominated by Brownian motion (small particles ${\lesssim}1\,\mu\mathrm{m}$), we obtain,
\begin{equation}
    \begin{aligned}
        & \tau_{\rm coag} \sim \frac{s_{\rm ice} \rho_{\bullet,\mathrm{ice}} }{ \sqrt{k_{\rm B} T/m_{\parti}} \rho_{\gas}} f_{\rm ice}^{-1}  \approx 47.5\, \mathrm{yr}\, \times \\
        & 
        \left(\frac{s_{\rm ice}}{1\mu m}\right)^\frac{5}{2} \left( 
        \frac{T}{135\mathrm{K}}\right)^{-\frac{1}{2}}
        \left( \frac{\rho_\bullet}{1\,\mathrm{g\,cm^{-3}}} \right)^\frac{3}{2}
        \left(\frac{\rho_{\gas}}{10^{-10} \si{g~cm^{-3}} }\right)^{-1} 
        \left(\frac{f_{\rm ice}}{0.01}\right)^{-1}
    \end{aligned}
\end{equation}
where $f_{\rm ice}$ is the ice-to-gas density ratio. In the above expression, we have inserted the values typical for the fiducial run.  In addition, we assumed ice grains have grown to $1\,\mu\mathrm{m}$, which is approximately the size where growth by Brownian motion is replaced by drift-induced mechanisms \citep{DullemondDominik2005,OrmelOkuzumi2013}. This time must be compared to the recycling timescale $\tau_{\rm rec}$. If the sublimation radius lies around or outside the Bondi radius, the outflow motion will have a velocity similar to the shear velocity, $v_{\rm rec} \sim \Omega R_{\rm subl}$. Therefore,
\begin{equation}
    \tau_{\rm rec} 
    \sim \frac{R_\mathrm{subl}}{v_\mathrm{rec}} \sim \Omega^{-1} \sim 1\,\mathrm{yr}
\end{equation}
As the coagulation timescale is much longer than the recycling timescale there will be no significant growth of ice grains during their transport away from the sublimation front. Therefore, pebble recycling will not be influenced by dust coagulation unless the coagulated grains have a very fractal structure (low $\rho_\bullet$; \citealt{Ormel2014}).

In contrast, for the vapour-dominated regime ice grains accumulate deep in the atmosphere (see \fg{inject_patterns}) and the coagulation timescale is much shorter. We have measured the coagulation timescale for the \texttt{pebble-5au} run at $t = 500\,\Omega^{-1}$,  when $f_{\rm ice} \approx 0.2$, $T \approx 200\, \mathrm{K}$ and $\rho_{\gas} \approx 10^{-10} \si{g~cm^{-3}}$ (see \fg{inject_5_au_profile}), as $\tau_{\rm coag} \sim 1\,\Omega^{-1}$. In addition, inside the critical streamline, where the streamlines are closed, the recyling timescale is given by the diffusion timescale $\tau_{\rm rec} \sim R_{\rm atm}^{2}/ D_{\gas} \sim 10^{4}\,\Omega^{-1}$. Since this coagulation timescale is much shorter than the recycling timescale, ice grains will grow to larger particles in the atmosphere and experience strong gravitational settling, causing them to fall back to the planet and the amount of grains that escape will be much lower. 
Instead of an extended ice grain distribution pervading the whole atmosphere seen in \fg{inject_patterns}, coagulation would limit the particles more to the sublimation front region. 
This would further exacerbate the growth of the planet's vapour content.

\subsection{Applications to disc chemistry}
In this work we have conducted local simulations, focusing on the consequences of sublimation for the putative planet atmosphere and its bulk composition. In the recycling-dominated regime the sublimated pebbles simply escape the simulation domain in the form of ice grains by instaneous freezing-out and there is no further interaction between the outflow vapour and the disc. However, the outflow vapour could possibly spread azimuthally through horseshoe orbits to fill the entire co-orbital region given the finite freeze-out rate of vapour \citep{MinissaleEtal2022}. 
Pebble recycling can therefore change the chemical abundance of disks locally.

In particular, the high accretion rates characterizing accreting giant planet would elevate ambient temperature significantly \citep{MachidaEtal2010,Szulagyi2017,Bergez-CasalouEtal2020}. $R_{\mathrm{subl}}$ could even expand to ${\sim}R_{\mathrm{Hill}}$, releasing vapour over a much wider region ($\sim$au) that is observationally accessible. 
Moreover, giant planets \citep{LinPapaloizou1979,Kley1999,KanagawaEtal2017} tend to open a deep gap, which allows the penetration of high flux UV-photons and promote further photo-chemistry \citep{BerginEtal2016,MiotelloEtal2019,BosmanEtal2021}.  \citet{CleevesEtal2015} proposed that the asymmetric features seen in molecular line emission might imply the existence of young giant planets. An example is the observed time-evolving SO asymmetry in HD 100546 \citep{BoothEtal2023}, which is proposed to trace the post shock region where the warm (${\sim}100\,\mathrm{K}$) environment caused by accreting giants is suitable for photo-chemistry to form SO. 

Quantitatively, these questions can best be addressed by global simulations. In a companion paper (Jiang et al.) we have implemented the sublimation module approach accordingly to address the potential of recycling of C-rich ices. This could modify the chemical abundance (e.g., C/O ratio) in the co-orbital region of the planet, which could result in distinct substructures as seen, for example, in the Molecules with ALMA at Planet-forming Scales survey (MAPS, \citet{OebergEtal2021}). Our phase change module can also be augmented with a chemical network to further understand the connection between young planets and their natal discs.

\section{Conclusions}
\label{sec:conclusion}
We have developed a phase change module based on the multi-fluid dust module in Athena++. This module treats the mass transfer, energy exchange and momentum conservation processes during sublimation and condensation (freezing-out) self-consistently, enabling us to study the coupled thermodynamic and hydrodyanmical effect of phase change properly in numerical simulations. 
In this work, we focus on investigating how ice sublimation influence the properties of the planetary atmosphere and the resulting vapour content through pebble accretion.

Our main findings are:
\begin{enumerate}
    \item The long-term evolutionary trend of the atmospheric vapour content is determined by the relative size of the sublimation radius $R_{\mathrm{subl}}$ and the atmosphere radius $R_{\mathrm{atm}}$. When $R_{\mathrm{subl}}$ exceeds $R_{\mathrm{atm}}$, incoming pebbles can be fully recycled and the vapour content reaches a steady value  -- the recycling-dominated limit. On the other hand, When $R_{\mathrm{subl}} < R_{\mathrm{atm}}$, vapour tends to be locked deep in the atmosphere and keep accumulating --  the vapour-dominated limit.
    \item The accumulation of vapour results in a positive feedback. $R_{\mathrm{atm}}$ expands due to material accumulation while $R_{\mathrm{subl}}$ shrinks due to the cooling caused by pebble sublimation. The divergence of $R_{\rm subl}$ and $R_{\rm atm}$ suppresses recycling and promotes further accumulation of vapour.
    \item Planets locating further out with respective to the star (lower ambient temperature) and higher pebble flux induced by either higher pebble-to-gas ratio or larger Stokes number tend to shrink the sublimation front, resulting in vapour-dominated regime. Conversely, with lower pebble fluxes or higher ambient temperatures, simulations lean more towards the recycling-dominated limit. However, the trend with pebble flux will probably reverse when frictional heating by pebbles is included (\se{caveats}).
    \item With increasing planet mass, $R_\mathrm{subl}$ increases faster than $R_\mathrm{atm}$. Consequently, the classification -- recycling, intermediate, or vapour-dominated -- tends to shift towards the former. Planet's atmospheres, while becoming bigger, may hold less (bound) vapour. 
    \item The intermediate regime is a transition between the recycling-dominated and the vapour-dominated cases. It features an instability triggered by composition gradients near the atmosphere radius and simulations tend to be characterized by weakened horseshoe orbits. 
    \item We develop estimators for $R_{\mathrm{subl}}$ and $R_{\mathrm{atm}}$ based on disc conditions and material properties. Applying these to an Earth-mass planet at 1 au in an MMSN disc, we find it to be recyling-dominatant in water vapour (i.e., a dry planet) while silicates are always in the vapour-dominated regime. Preventing carbonaceous material to be accumulated, requires the ambient temperature to be $\gtrsim$200\,K.
\end{enumerate}

The phase change module developed by us is general. In the future, it can be applied to address questions regarding the chemical compositions influenced by the planet-disc interaction and planetesimal formation near the snowline, among others.

\section*{Acknowledgements}
The authors thank the referee for their positive assessment of this manuscript. The authors appreciate fruitful discussions with Zhaohuan Zhu, Allona Vazan, Xuening Bai, Zhuo Chen, Kees Dullemond, and Mario Flock. The authors acknowledge the Tsinghua Astrophysics High-Performance Computing platform at Tsinghua University for providing computational and data storage resources that have contributed to the research results reported within this paper. 
CWO acknowledges support by the National Natural Science Foundation of China (grant no. 12250610189). 
RK acknowledges financial support via the Heisenberg Research Grant funded by the German Research Foundation (DFG) under grant No. KU2849/9.
\section*{Data Availability}
The data underlying this article will be shared on reasonable request to the corresponding author.
 





\bibliographystyle{mnras}
\bibliography{/media/yu/Data/Documents/Refs/Bibs/ads} 



\appendix

\section{Gas-only run}
\label{sec:gas_only}
\begin{figure*}
    \includegraphics[width=2\columnwidth]{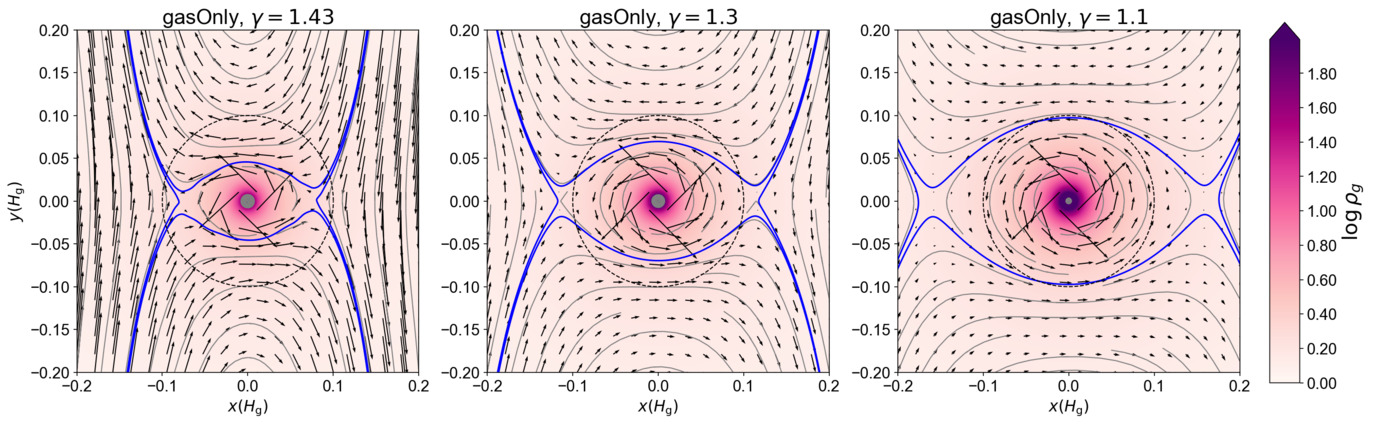}
    \caption{Steady state flow patterns of \text{gasOnly} case with different adiabatic index $\gamma$. The color denotes gas density in log scale. Their x and y axies are all plotted in the unit of local scale height. The black dashed line in each plot denotes Bondi radius (\eq{bondi_radius}) and the blue solid line is ``critical streamline'' (\se{coupled}).}
    \label{fig:gasonly_pattern}
\end{figure*}

In this part, we consider the situation where an inviscid gas flow passes through a planet atmosphere, a problem which has already been studied before \citep{OrmelEtal2015,FungEtal2015,KurokawaTanigawa2018,BethuneRafikov2019}. The parameters of these \texttt{gasOnly} runs are listed in \tb{simu_paras}. By design, the gas is adiabatic and $\gamma = 1.43$ (assume $71\%$ $\mathrm{H_{2}}$ and $29\%$ He). We also test cases with $\gamma = 1.3, 1.1$ (other parameters are the same), see \fg{gasonly_pattern}. The first three panels show steady state flow patterns of cases with $\gamma = 1.43,1.3$ and $1.1$ respectively, where the color denotes gas density (on logarithmic scale).

Overall the flow patterns show the typical horseshoe orbit and inner rotational atmosphere shown in previous studies.
With the decrease of the adiabatic index, the atmosphere expands and rotational support becomes more and more important, which is clearly shown in the upper panel of \fg{gasonly_profile}. Also, the density increases significantly. As $\gamma$ become smaller, pressure support becomes less effective, resulting in more gas falling in, which is also seen when comparing the adiabatic and isothermal atmospheres in \citet{BethuneRafikov2019}. 

For an adiabatic EOS, the entropy should be conserved, which is approximately true for high $\gamma$ runs. However, for the run with $\gamma = 1.1$, the entropy drops steeply to nearly 0 at the inner boundary, concommitant with the abrupt increase of gas density. We verified that no energy flux was flowing through the inner boundary consistent with the reflective boundary condition. We suspect that the unphysical entropy violation is caused by the large azimuthal velocity that reaches Keplerian values near the inner boundary, which potentially induces significant numerical viscosity. We verified that mitigating velocity gradients by adopting gravitational softening \citep{FungEtal2019,ZhuEtal2021} or a larger value of the inner radius or higher resolution, suppresses the artificial entropy variation, while solving for the entropy equation instead of the energy equation \citep{MignoneEtal2007,MignoneEtal2012} eliminated the issue. In the simulations runs in the main text we simply monitor the entropy evolution, and re-run simulations at higher resolution when the entropy variation exceeds ${\sim}10\%$ and begins to alter the flow pattern.  

\begin{figure}
    \includegraphics[width=\columnwidth]{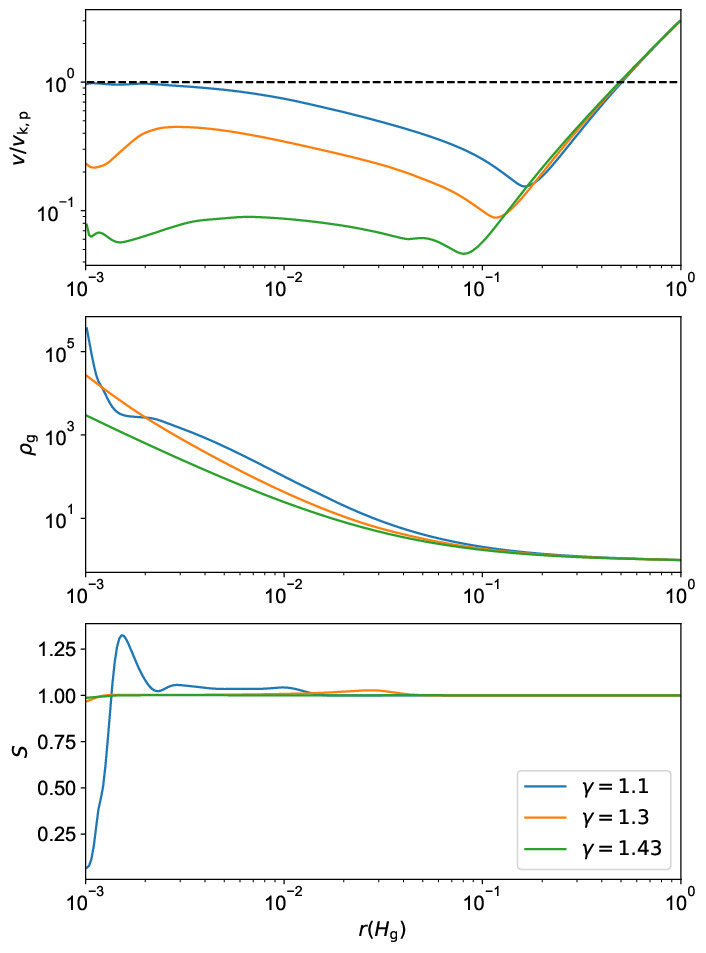}
    \caption{Comparison of azimuthally averaged radial density and velocity profiles of run \texttt{gasOnly}. The velocity profile is in unit of the planet's Keplerian velocity $v_{\mathrm{k}} = \sqrt{m/r}$ and the density profile is normalized by the disc's unperturbed value $\rho_{0}$.}
    \label{fig:gasonly_profile}
\end{figure}

\section{Estimation of sublimation front}
\label{sec:estimate}
Combining \eqs{R_subl_quanti}{adiabatic_solution},
\begin{equation}
    \label{eq:scaling_1}
    \begin{aligned}
        &f_{\rm vapr,peak} \left[1+\frac{\gamma-1}{\gamma} R_{\rm Bondi}\left(\frac{1}{R_{\rm subl}}-\frac{1}{r_{\rm max}}\right) \right]^{\frac{\gamma}{\gamma-1}} P_{\rm disc}\\
        & = P_{\rm eq,0} \exp \left(\frac{-T_{\rm a}}{T_{\mathrm{subl}}^{\prime}} \right) \\
        & \Longrightarrow \log \left(f_{\rm ice} \frac{P_{\rm disc}}{{P_{\rm eq,0}}}\right) + \frac{\gamma}{\gamma-1} \log \chi = -\frac{T_{\rm a}}{T_{\mathrm{subl}}^{\prime}}
    \end{aligned}
\end{equation}
where $\chi = 1+\frac{\gamma-1}{\gamma}R_{\rm Bondi}\left(\frac{1}{R_{\rm subl}}-\frac{1}{r_{\rm max}}\right)$. $T_{\mathrm{subl}}^{\prime}$ is the modified temperature at the sublimation front, which takes into accout the sublimation cooling compared to $T_{\mathrm{subl}}$ in \eq{adiabatic_solution}. Assuming all the ice sublimates at $R_{\mathrm{subl}}$ (In reality sublimation happens in a finite band, see \fg{inject_fiducial_fraction}), latent heat should be deducted from the gas internal energy,
\begin{equation}
    \frac{P_{\gas}}{\rho_{\gas} (\gamma-1)} - L_{\mathrm{heat}} f_{\rm vap,peak} = \frac{k_{\mathrm{B}} T_{\mathrm{subl}}^{\prime} }{\mu_{\gas}m_{\mathrm{p}} (\gamma-1) }
\end{equation}
For $P_{\gas}$ and $\rho_{\gas}$ we still take the adiabatic solution, which is a very rough zero-order approximation. Then $T_{\mathrm{subl}}^{\prime}$ yields,
\begin{equation}
    \label{eq:scaling_2}
    T_{\mathrm{subl}}^{\prime} = T_{\mathrm{subl}} - \frac{L_{\mathrm{heat}} f_{\mathrm{vap,peak}} (\gamma-1)}{k_{\mathrm{B}}/(\mu_{\gas} m_{\mathrm{p}})}
\end{equation}
Combining \eqs{scaling_1}{scaling_2}, $R_{\mathrm{subl}}$ can be numerically solved.


\bsp	
\label{lastpage}
\end{document}